\documentclass[preprints,article,submit,accept,moreauthors]{Definitions/mdpi}

\firstpage{1} 
\pubvolume{1}
\issuenum{1}
\articlenumber{0}
\pubyear{2023}
\copyrightyear{2023}
\datereceived{ } 
\daterevised{ } 
\dateaccepted{ } 
\datepublished{ } 
\hreflink{https://doi.org/} 

\Title{On Finding Optimal (Dynamic) Arborescences}

\TitleCitation{On Finding (Dynamic) Optimal Arborescences}
 
\Author{Joaquim Espada $^{1,2}$, Alexandre P. Francisco$^{1,2}$, Tatiana Rocher$^{1}$, Lu\'is M. S. Russo$^{1,2}$ and C\'atia Vaz$^{2,3}$}

\AuthorNames{Joaquim Espada, Tatiana Rocher, Lu\'is M. S. Russo, Alexandre P. Francisco* and C\'atia Vaz}

\AuthorCitation{Espada, J.; Francisco, A. P.; Rocher, T.; Russo, L. M. S.,; Vaz, C.}
\address{%
$^{1}$ \quad Instituto Superior Técnico, Universidade de Lisboa, Portugal\\
$^{2}$ \quad Instituto de Engenharia de Sistemas e Computadores, I\&D em Lisboa (INESC-ID Lisboa), Portugal\\
$^{3}$ \quad Instituto Superior de Engenharia de Lisboa, Instituto Politécnico de Lisboa, Portugal\\
}

\usepackage[usestackEOL]{stackengine}
\usepackage{mdframed}
\usepackage{pgfpages}
\usepackage{pgfkeys}
\usepackage{algpseudocode}
\usepackage{algorithm}
\usepackage{subcaption}
\usepackage{tikz}
\usetikzlibrary{arrows,arrows.meta,shapes,positioning,quotes,matrix,backgrounds}
\tikzset{vertex/.style = {shape=circle,draw,minimum size=1.5em}}
\tikzset{edge/.style = {->,> = latex}}
\tikzstyle{vertex}=[circle,fill=black!25,minimum size=20pt,inner sep=0pt]
\newcommand{\codelineabove}{\noindent\rule{\linewidth}{0.1pt}\par}
\newcommand{\codelinebelow}{\par\noindent\rule{\linewidth}{0.1pt}}
\newcommand{\node}{\mathit{node}}
\newcommand{\nodeF}{\mathit{nodeF}}

\newcommand{\child}{\mathit{edge}}
\newcommand{\queues}{\mathit{queues}}
\newcommand{\map}{\mathit{map}}
\newcommand{\cycle}{\mathit{cycle}}
\newcommand{\compare}{\mathit{compare}}
\newcommand{\roots}{\mathit{roots}}
\newcommand{\varmax}{\mathit{max}}
\newcommand{\rset}{\mathit{rset}}
\newcommand{\inEdgeNode}{\mathit{inEdgeNode}}

\newcommand{\cycleEdgeNode}{\mathit{cycleEdgeNode}}
\newcommand{\minNodeF}{\mathit{minNodeF}}

\newcommand{\rep}{\mathit{rep}}
\newcommand{\reduct}{\mathit{cost}}

\newcommand{\POP}{\texttt{POP}}
\newcommand{\MELD}{\texttt{MELD}}

\newcommand{\EDGE}{\texttt{EDGE}}
\newcommand{\INIT}{\texttt{INIT}}
\newcommand{\WMAKE}{\texttt{WMAKE-SET}}
\newcommand{\SMAKE}{\texttt{SMAKE-SET}}
\newcommand{\EXTRACT}{\texttt{EXTRACT-MIN}}
\newcommand{\SFIND}{\texttt{SFIND}}
\newcommand{\WFIND}{\texttt{WFIND}}
\newcommand{\SADD}{\texttt{SADD-WEIGHT}}
\newcommand{\SUNION}{\texttt{SUNION}}
\newcommand{\WUNION}{\texttt{WUNION}}
\newcommand{\PARENT}{\texttt{PARENT}}
\newcommand{\CHILDREN}{\texttt{CHILDREN}}

\newcommand{\CONTINUE}{\texttt{continue}}
\newcommand{\BREAK}{\texttt{break}}

\renewcommand{\contentleftcolumn}{}

\abstract{Let $G=(V,E)$ be a directed and weighted graph with vertex set $V$ of size $n$ and edge set $E$ of size $m$, such that each edge $(u,v)\in E$ has a real-valued weight $w(u,c)$.
An arborescence in $G$ is a subgraph $T=(V,E')$ such that for a vertex $u\in V$, the root, there is a unique path in $T$ from $u$ to any other vertex $v\in V$.
The weight of $T$ is the sum of the weights of its edges.
In this paper, given $G$, we are interested in finding an arborescence in $G$ with minimum weight, i.e., an optimal arborescence.
Furthermore, when $G$ is subject to changes, namely edge insertions and deletions, we are interested in efficiently maintaining a dynamic arborescence in $G$.
This is a well known problem with applications in several domains such as network design optimization and in phylogenetic inference.
In this paper we revisit algorithmic ideas proposed by several authors for this problem, we provide detailed pseudo-code as well as implementation details, and we present experimental results on large scale-free networks and on phylogenetic inference.
Our implementation is publicly available at \url{https://gitlab.com/espadas/optimal-arborescences}.}


\begin{document}

\section{Introduction}
The problem of finding an optimal arborescence in directed and weighted graphs is one of the fundamental problems in graph theory with several practical applications.
It has been found in modeling broadcasting~\cite{li2006construction}, network design optimization~\cite{fortz2018optimal},  subroutines to approximate other problems, such as the traveling salesman problem~\cite{gerhard1994traveling}, and it is also closely related to the Steiner problem~\cite{cong1998efficient}.
Arborescences are also found in multiple clustering problems, from taxonomy to handwriting recognition and image segmentation~\cite{coscia2018using}.
In phylogenetics, optimal arborescences are useful representations of probable phylogenetic trees~\cite{zhou2018grapetree,vaz2021distance}.

Chu and Liu~\cite{chu1965shortest}, Edmonds~\cite{edmonds1967optimum}, and Bock~\cite{bock1971algorithm} proposed independently a polynomial time algorithm for the static version of this problem.
The algorithm by Edmonds relies on a contraction phase followed by an expansion phase.
A faster version of Edmonds algorithm was proposed by Tarjan~\cite{tarjan1977finding}, running in $O(m\log n)$ time.
Camerini et al.~\cite{camerini1979note} corrected the algorithm proposed by Tarjan, namely the expansion procedure.
The fastest known algorithm was proposed later by Gabow et al.~\cite{gabow1986efficient}, with improvements in the contraction phase and running in $O(n\log n + m)$ time. 
Fischetti and Toth~\cite{fischetti1993efficient} also address this problem restricted to complete directed graphs, relying on the Edmonds algorithm.
The algorithms proposed by Tarjan, Camerini et al., and Gabow et al., rely on elaborated constructions and advanced data structures, namely for efficiently keeping mergeable heaps and disjoint sets.

As stated by Aho et al.~\cite{aho1997emerging}, ``efforts must be made to ensure that promising algorithms discovered by the theory community are implemented, tested and refined to the point where they can be usefully applied in practice.''
The transference of algorithmic ideas and results from algorithm theory to practical applications can be however considerable, in particular when dealing with elaborated constructions and data structures, a well known challenge in algorithm engineering~\cite{sanders2009algorithm}.

Although there are practical implementations of the Edmonds algorithm, such as the implementation by Tofigh and Sj\"olund~\cite{Tofigh} or the implementation in NetworkX~\cite{Hagberg}, most of them neglect these elaborated constructions. Even though the Tarjan version is mentioned in the implementation by Tofigh and Sj\"olund, they state in the source code that its implementation is left to be done.
An experimental evaluation is also not provided together with most of these implementations.
Only recently Espada~\cite{espada2019} and B{\"o}ther et al.~\cite{bother2023efficiently} provided and tested efficient implementations, taking into account more elaborated constructions.
We highlight in particular the experimental evaluation by B{\"o}ther et al. with respect to the use of different mergeable heap implementations, and their conclusions pointing out that the Tarjan version is the most competitive in practice.

These experimental results are for the static version.
As far as we know, only Pollatos, Telelis and Zissimopoulos~\cite{zissimopoulosfully,pollatos2006updating} studied the dynamic version of the problem of finding optimal arborecences.
Although Pollatos et al. provided experimental results, they did not provide implementation details nor, as far as we known, a publicly available implementation.
Their results point out that the dynamic algorithm is particularly interesting for sparse graphs, as is the case of most real networks, which are in general scale-free graphs~\cite{barabasi2016}.

In this paper we present detailed pseudo-code and a practical implementation of Edmonds algorithm taking into account the construction by Tarjan~\cite{tarjan1977finding} and the correction by Camerini~\cite{camerini1979note}.
Based on this implementation, and on the ideas by Pollatos et al., we present also an implementation for the dynamic version of the problem.
As far as we know this is the first practical and publicly available implementation for dynamic directed and weighted graphs using this construction. Moreover, we provide generic implementations in the sense that 
a generic comparator is given as parameter and, hence, we are not restricted to weighted
graphs; we can find the optimal arborescence on any graph equipped with a total order on the set of edges.
We provide also experimental results of our implementation for large scale-free networks and in phylogenetic inference use cases, detailing design choices and the impact of used data structures.
Our implementation is publicly available at  \url{https://gitlab.com/espadas/optimal-arborescences}.

The rest of the paper is organized as follows. In Section~\ref{sec:prob} we introduce the problem of finding optimal arborescences, and we describe both Edmonds algorithm and the Tarjan algorithm, including the correction by Camerini et al.
In Section~\ref{sec:dynarbo} we present the dynamic version of the problem and the studied algorithm.
We provide implementation details and data structures design choices in Section~\ref{sec:impl}.
Finally, we present and discuss experimental results in Section~\ref{sec:res}.

\section{Optimal arborescences}
\label{sec:prob}

Both Edmonds and Tarjan algorithms proceed in two phases: a contraction phase followed by an expansion phase.
The {\em contraction phase} then maintains a set of candidate edges for the optimal arborescence under construction.
This set is empty in the beginning.
As this phase proceeds, edges selected may form cycles which are contracted to form super-vertices.
The {\em contraction phase} ends when no contraction is possible and all vertices have been processed.
In the {\em expansion phase} super-vertices are expended in reverse order of their contraction,
and one edge is discarded per cycle to form the arborescence of the original graph.
The main difference between both algorithms is on the contraction phase. 

\subsection{Edmonds algorithm}\label{algo:edmonds}

Let $G=(V,E)$ be a directed and weighted graph with vertex set $V$ of size $n$ and edge set $E$ of size $m$, such that each edge $(u,v)\in E$ has a real-valued weight $w(u,v)$.
Let each contraction mark the end of an iteration of the algorithm and, for iteration $i$, let $G_i$ denote the graph at that iteration, $D_i$ be the set of selected vertices in iteration $i$, $E_i$ be the set of selected edges incident on the selected vertices, and $Q_i$ be the cycle formed by the edges in $E_i$, if any.

The algorithm starts with $G_0 = G$, $D_i$ and $E_i$ initialized as empty sets, and $Q_i$ initialized as an empty graph, for all iterations $i$.

\subsubsection{Contraction phase}
The algorithm proceeds by selecting vertices in $G_i$ which are not yet in $D_i$. If such a vertex exists, then it is added to $D_i$ and the minimum weight incident edge on it is added to $E_i$.
The algorithm stops if either a cycle is formed in $E_i$ or if all vertices of $G_i$ are in $D_i$.

If $E_i$ holds a cycle, then we add the edges forming the cycle to $Q_i$ and we build a new graph $G_{i+1}$ from $G_i$, where the vertices in the cycle are contracted into a single super-vertex $v^{i+1}$.
Edges $(u, v) \in G_i$ are added to $G_{i+1}$ and updated as follows.
\begin{enumerate}
  \item Loop edge removal: if $u, v \in Q_i$, then $(u, v) \notin G_{i+1}$.
  \item Unmodified edge preservation: if $u, v \notin Q_i$ then $(u, v) \in G_{i+1}$.
  \item Edges originating from the new vertex: if $u \in Q_i \land v, \notin Q_i$, then $(v^{i+1}, v) \in G_{i+1}$.
  \item Edges incident to the new vertex: if $u \notin Q_i \land v \in Q_i$, then $(u, v^{i+1}) \in G_{i+1}$
  and $w(u, v^{i+1}) = w(u, v) + \sigma_{Q_i} - w(u', v)$.
\end{enumerate}
Here $w(u, v^{i+1})$ is the weight of the edge $(u, v^{i+1})$ in $G_{i+1}$,
$w(u, v)$ is the weight of $(u, v)$ in $G_i$,
$\sigma_{Q_i}$ denotes the maximum edge weight in the cycle $Q_i$,
and $w(u', v)$ is the weight of the edge in the cycle incident to vertex $v$.
After the weights are updated, the algorithm continues the contraction phase with the next iteration $i+1$.

The contraction phase ends when there are no more vertices to be selected in $G_i$, for some iteration $i$.

\subsubsection{Expansion phase}
The final content of $E_i$ holds an arborescence for the graph $G_i$.
Let $H$ denote a subgraph formed by the edges of $E_i$.
For every contracted cycle $Q_i$, we add to $H$ all cycle edges except one.
If the contracted cycle is a root of $H$, we discard the cycle edge with the maximum weight.
If the contracted cycle if not a root of $H$, we discard the cycle edge that shares the same destination as an edge currently in $H$.
The algorithm proceeds in reverse with respect to the contraction phase, examining graph $G_{i-1}$ and the cycle $Q_{i-1}$.
This process continues until all contractions are undone, and $H$ is the final arborescence.

\subsubsection{Illustrative example}
\begin{figure}[H]
	\centering
	\subcaptionbox{Input weighted directed graph.\label{fig:input_graph_edmonds}}{\includegraphics{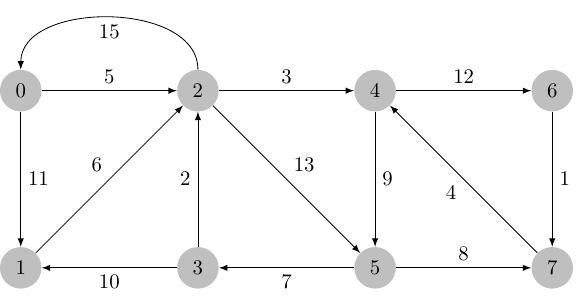}}
	\subcaptionbox{Minimum weight incident edges in every vertex of graph $G_0$, colored in red.\label{fig:g0_min_edges}}{\includegraphics{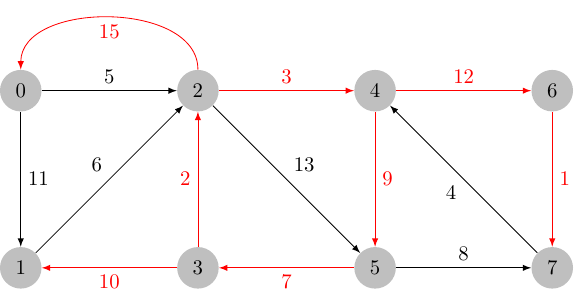}}
	\subcaptionbox{Cycle in $G_0$ colored in green.\label{fig:g0_cycle}}{\includegraphics{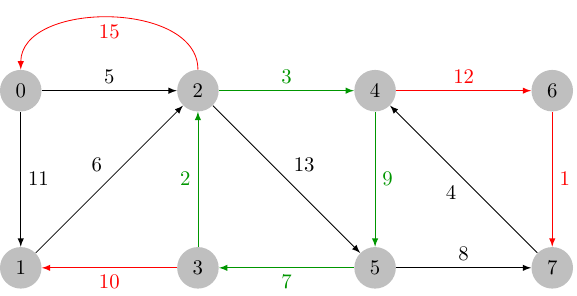}}
	\caption{Identification of the cycle $Q_0$ in $G_0$.}
\end{figure}
Let us consider the graph in Figure~\ref{fig:input_graph_edmonds}. Let the input graph be denoted by $G_{0}$. In Figure~\ref{fig:g0_min_edges}, the minimum weight incident edges in every vertex of $G_0$ are colored in red, those edges are added $E_0 = \{(2,0),(3,1), (3,2), (2,4), (5,3),(4,6), (6,7) \}$. The green edges in Figure~\ref{fig:g0_cycle} form a cycle and are added to $Q_0 = \{(3,2),(2,4),(4,5),(5,3)\}$. The cycle $Q_0$ must be then contracted. The maximum weight edge in $Q_0$ is the edge $(4,5)$ and $\sigma_{Q_0} = 9$.

\begin{figure}[H]
	\centering
	\subcaptionbox{Contracted version of graph $G_{0}$ named $G_1$.\label{fig:g1}}{\includegraphics{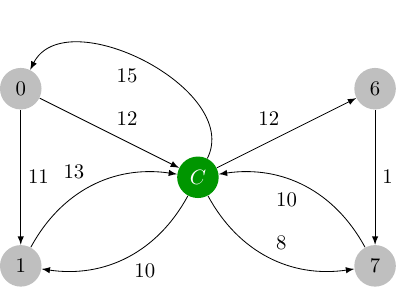}}
    \hfill
	\subcaptionbox{Minimum weight incident edges in every vertex of graph $G_1$.\label{fig:g1_min_edges}}{\includegraphics{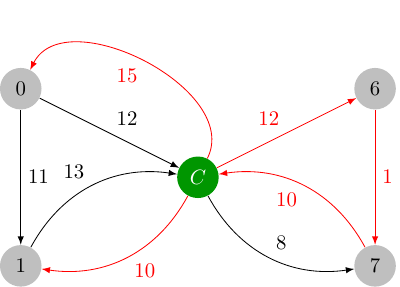}}
	\subcaptionbox{Cycle $Q_1$ in graph $G_1$ colored in green.\label{fig:g1_cycle}}{\includegraphics{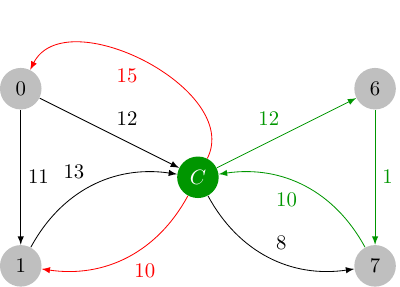}}
	\caption{Contraction of cycle  $Q_{0}$ and identification of cycle $Q_1$.}
\end{figure}

In Figure~\ref{fig:g1} it is shown the contracted version of $G_{0}$, named $G_1$, with the reduced costs already computed. In Figure~\ref{fig:g1_min_edges}, the minimum weight incident edges are highlighted in red and they are added to $E_1 = \{(C,0),(C,1), (C,6), (7,C), (6,7)\}$. Still in Figure~\ref{fig:g1_cycle}, cycle $Q_1 = \{(C,6),(6,7),(7,C)\}$ is marked in green and it will be contracted.
The maximum weight edge of $Q_1$ is edge $(C,6)$ and $\sigma_{Q_1} = 12$.

\begin{figure}[H]
	\centering
	\subcaptionbox{Contracted version of $G_{1}$ named $G_2$.\label{fig:g2}}{\includegraphics{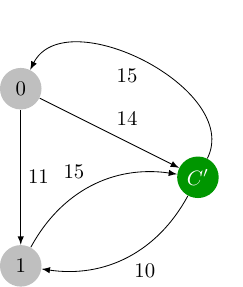}}
 \hfill
	\subcaptionbox{Minimum weight incident edges in $G_2$.\label{fig:g2_min_edges}}{\includegraphics{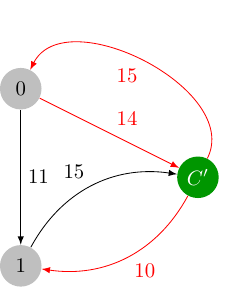}}
 \hfill
	\subcaptionbox{Cycle in $G_2$ colored in green.\label{fig:g2_cycle}}{\includegraphics{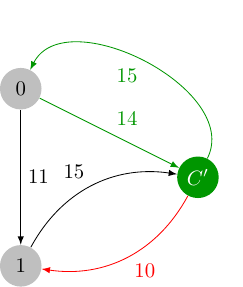}}
\caption{Contraction of cycle $Q_1$ and identification of cycle $Q_2$.}
\end{figure}

The contracted version of $G_1$, named $G_2$, is presented in Figure~\ref{fig:g2}. Minimum weight incident edges in every vertex of $G_2$ is marked in red in Figure~\ref{fig:g2_min_edges}, and they are added to $E_2 = \{(0,C'), (C',0), (C',1)\}$. $E_2$ has a cycle, $Q_2=\{(0,C'),(C',0)\}$, that is colored in green in Figure~\ref{fig:g2_cycle}.
This cycle must be also contracted. The maximum weight edge of $Q_2$ is edge $(0,C')$ and $\sigma_{Q_2} = 15$.

\begin{figure}[H]
	\centering
		\centering
		\subcaptionbox{Contracted version of $G_{2}$ named $G_3$.\label{fig:g3}}{\includegraphics{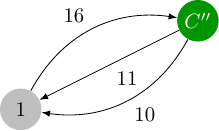}}
  \hfill
		\subcaptionbox{Minimum weight incident edges in $G_3$.\label{fig:g3_min_edges}}{\includegraphics{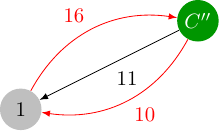}}
  \hfill
		\subcaptionbox{Cycle $Q_3$ in $G_3$ colored in green.\label{fig:g3_cycle}}{\includegraphics{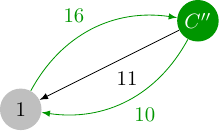}}
  \hfill
		\subcaptionbox{$G_4$.\label{fig:g4}}{\includegraphics{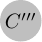}}
		\caption{Contraction of $Q_2$, identification of $Q_3$, and final graph $G_4$ after $Q_3$ contraction.}
\end{figure}

Since $E_2$ contains a cycle, a contraction is required and we obtain the graph $G_3$ in Figure~\ref{fig:g3}. The minimum incident edges in $G_3$ are marked in red in Figure~\ref{fig:g3_min_edges}, and added to $E_3 =  \{(1,C''), (C'',1)\}$. Note that $E_3$ contains another cycle $Q_3 = \{(1,C''), (C'',1)\}$. The maximum weight edge present in $Q_3$ is $(1,C'')$ and $\sigma_{Q_3} = 16$. A final contraction of cycle $Q_3$ is required, leading to $G_4$ with a single vertex, Figure~\ref{fig:g4}. In this last iteration we have $E_4 = \emptyset$ and $Q_4 = \emptyset$, ending the contraction phase. 

\begin{figure}[H]
	\centering
	\subcaptionbox{Expansion subgraph $H$ in iteration $i = 3$.\label{fig:sol_0}}{\includegraphics{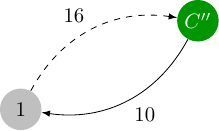}}
	\hfill
	\subcaptionbox{Expansion subgraph $H$ in iteration $i = 2$.\label{fig:sol_1}}{\includegraphics{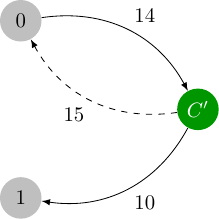}}
	\hfill
	\subcaptionbox{Expansion subgraph $H$ in iteration $i = 1$.\label{fig:sol_2}}{\includegraphics{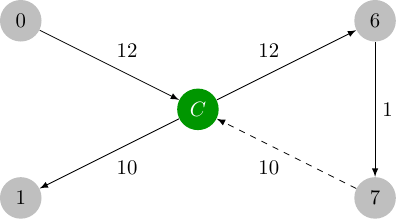}}
	
	\vspace{5mm}
	\subcaptionbox{Expansion subgraph $H$ in iteration $i = 0$.\label{fig:sol_3}}{\includegraphics{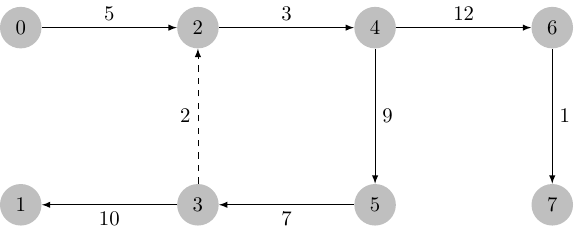}}
	\caption{Expansion phase and optimal arborescence.}
\end{figure}

The expansion phase expands the cycles formed in reverse order kicking out one edge per cycle. The removed edges are presented as dashed edges. Let $H = E_4$ and decrement $i$. Vertex $C'''$ is a root of $H$ since there is no edge directed towards $C'''$. In this case every edge of $Q_3$ is added to $H$ except the maximum weight edge of the cycle as shown in Figure~\ref{fig:sol_0}.
In iteration $i=2$, note that $H = \{(C',1)\}$, and vertex $C'$ is a root, therefore every edge from $Q_2$ must be added to $H$ except the maximum weight edge $(C',0)$.
In the next iteration, $i = 1$, vertex $C$ is not a root of $H$ since edge $(0,C) \in H$. In this case, we add every edge from $Q_1$ except the ones that share the destination with edges in $H$, as illustrated in Figure~\ref{fig:sol_2}.
Regarding the final expansion, $H = \{(0,C),(C,1),(C,6),(6,1)\}$ implies that $C$ is not a root. Every edge in $Q_0 = \{(3,2),(2,4),(4,5),(5,3)\}$ except edge $(3,2)$ is added to $H$. The optimal arborescence of $G_0$ is shown in Figure~\ref{fig:sol_3}.

\subsection{Tarjan algorithm}\label{sec:tarjan}
The algorithm proposed by Tarjan~\cite{tarjan1977finding} is built on Edmonds algorithm, but it relies on advanced data structures to become more efficient, namely in the contraction phase.
The algorithm builds in this phase a subgraph $H=(V,E')$ of $G=(V,E)$ such that $H$ contains the selected edges.
The optimum arborescence could then be extracted from $H$ through a depth-first search, taking into account Lemma 2 in Tarjan paper~\cite{tarjan1977finding}.
This lemma states that there is always a simple path in $H$ from any vertex $u$ in a root strongly connected component $S$ to any vertex $v$ in the weakly connected component containing $S$.
Camerini et al.~\cite{camerini1979note} provided however a counter-example for this construction, and they proposed a correction that relies on a auxiliary forest $F$ and that we discuss below.

The algorithm by Tarjan keeps track of weakly and strongly connected components in $G$, as well as non examined edges entering each strongly connected component.
The bookkeeping mechanism used the union-find data structure~\cite{galler1964improved} to maintain disjoint sets.
Let \texttt{SFIND}, \texttt{SUNION}, \texttt{SMAKE-SET} denote operations on strongly connected components and \texttt{WMAKE-SET}, \texttt{WFIND}, \texttt{WUNION} on weakly connected components. Find operations allow to find the component where a given vertex lies in, union operations allow to merge two components together, and make-set operations allow to initialize singleton components for each vertex.
Non examined edges are kept through a collection of priority queues, implemented as mergeable heaps.
Let \texttt{MELD}, \texttt{EXTRACT-MIN}, and \texttt{INIT} denote the operations on heaps, where the meld operation allow to merge two heaps, the extract-min allows to get and remove the minimum weight element, and the initialization operation allows to initialize a heap from a list of elements. We consider also the \texttt{SADD-WEIGHT} operation that allows to add a constant weight to all edges incident on a given strongly connected component in constant time.
Note that edges incident on given strongly connected component are maintained in a priority queue where they are compared taking into account its weight and the constant weight added to that strongly connected component.

The correction proposed by Camarini et al. requires us then to maintain a forest $F$ and a set \texttt{rset} that holds the roots of the optimal arborescence, i.e., the vertices without incident edges. 
Each node of forest $F$ has associated an edge of $G$, a parent node, and a list of children.

\subsubsection{Initialization}
Data structures are initialized as follows. $\queues$ is an array of heaps, initialized with an heap for each vertex $v$ containing incident edges on $v$. $\roots$ is the list of vertices to be processed, initialized as $V$.
The forest $F$ is initialized as empty as well as the set $rset$. 
Four auxiliary arrays are also needed to build $F$ and the optimal arborescence, namely $\inEdgeNode$ that for each vertex $v$ stores a node of $F$ associated with the minimum weight edge incident in $v$, $\pi$ that stores the leaf nodes of $F$, $\cycleEdgeNode$ that stores for each cycle representative vertex $v$ the list of cycle edge nodes in $F$, and $\varmax$ that stores for each strongly connected component the target of the maximum weight edge.
These arrays are initialized as follows.

\codelineabove
\begin{algorithmic}
\State $roots \leftarrow \emptyset$ \Comment {Set of vertices to process.}
\For {each $v \in V$} 
  \State $\queues[v] \leftarrow \INIT(v, L[v])$ \Comment{$L[v]$ refers to the list of edges incident in $v$.}
  \State $\SMAKE(v)$, $\WMAKE(v)$
  \State $\roots \leftarrow \roots \cup \{v\}$ 
  \State $\varmax[v] \leftarrow v$
  \State $\inEdgeNode[v] \leftarrow \mathit{null}$ 
  \State $\pi[v] \leftarrow \mathit{null}$
  \State $\cycleEdgeNode[v] \leftarrow \emptyset$
\EndFor
\State $F \leftarrow \emptyset$
\State $\rset \leftarrow \emptyset$
\end{algorithmic} 
\codelinebelow

\subsubsection{Contraction phase}
The contraction phase proceeds while $\roots\neq\emptyset$ as follows. It pops a vertex $r$ from $\roots$ and
it verifies if there are incident edges in $r$ such that they do not belong to a contracted strongly connected component.
If there are such edges, then it extracts the one with minimum weight; otherwise it stops and it continues with another
vertex in $\roots$. The pseudo-code is as follows.

\codelineabove
\begin{algorithmic}
  \State $r \leftarrow \POP(roots)$
  \If {$\queues[r] \neq \emptyset$}
    \State $(u, r) \leftarrow \EXTRACT(\queues[r])$
    \While {$\queues[r] \neq \emptyset$ and $\SFIND(u) = \SFIND(r)$}
        \State $(u, r) \leftarrow \EXTRACT(\queues[r])$
    \EndWhile
    \If {$\SFIND(u) = \SFIND(r)$}
        \State $\rset \leftarrow \rset \cup \{r\}$
        \State $\CONTINUE$
    \EndIf 
  \Else
    \State $\rset \leftarrow \rset \cup \{r\}$
    \State $\CONTINUE$ 
  \EndIf
\end{algorithmic}
\codelinebelow

Once an incident edge on $r$ is found that does not lie within a strongly connected component, i.e., that is incident on a contracted strongly connected component, we must update forest $F$.
Hence we create a new node $\minNodeF$ in forest $F$ associated with edge $(u,r)$.
If $r$ is not part of a strongly connected component, i.e., $r$ is not part of a cycle, then $\minNodeF$ becomes a leaf of $F$. Otherwise, $F$ must be updated by making $\minNodeF$ a parent of the trees of $F$ that are part of the strongly connected component. The following pseudo-code details this updating of forest $F$.

\codelineabove
\begin{algorithmic}
  \State Create the node $\minNodeF$ in forest $F$ associated with the edge $(u, r)$
  \If {$\cycleEdgeNode[r] = \emptyset$}
    \State $\pi[r] \leftarrow \minNodeF$
  \Else
    \For {each $n \in \cycleEdgeNode[r]$}
      \State $\PARENT(n) \leftarrow \minNodeF$
      \State $\CHILDREN(\minNodeF)\leftarrow \CHILDREN(\minNodeF) \cup \{n\}$
    \EndFor
  \EndIf
\end{algorithmic}
\codelinebelow

The next step is to verify if $(u,r)$ forms a cycle with minimum weight edges formerly selected.
It is enough to check if $(u,r)$ connects vertices in the same weakly connected components.
Note that $(u,r)$ is incident on a root and, if $u$ lies in the same weakly connected component as $r$, then adding $(u,r)$ forms necessarily a cycle.
Assuming that adding $(u,r)$ does not form a cycle, we perform the union of the sets representing the two weakly connected components to which $u$ and $r$ belong, i.e., $\WUNION(u,r)$. We update also the $\inEdgeNode[r]$ array as $r$ now has an incident edge selected.

If adding $(u,r)$ forms a cycle a contraction is performed. The contraction procedure starts firstly by finding the edges involved in the cycle, using a backward depth-first search. During this process, a $map$ is initialized where the edge is associated to its $F$ node (the map key).
Then the maximum weight edge in the cycle is found, the reduced costs are computed and the weight of the edges is updated. 
Note that the min-heap property is always maintained when reducing the costs without running any kind of procedure to ensure it, since the constant $reduced$ is added to every edge in a given priority queue.
Arrays $\inEdgeNode$ and $\cycleEdgeNode$ are updated, and heaps involved in the cycle are merged. The pseudo-code is as follows.

\codelineabove
\begin{algorithmic}
  \If {$\WFIND(u) \neq \WFIND(r)$}
    \State $\inEdgeNode[r] \leftarrow \minNodeF$
    \State $\WUNION(u, r)$
  \Else
    \State $\inEdgeNode[r] \leftarrow \mathit{null}$
    \State $\cycle \leftarrow \{\minNodeF\}$
    \State Let $\map$ denote a map.
    \State $\map[\minNodeF] \leftarrow (u, r)$
    \State $u \leftarrow \SFIND(u)$
    \While {$\inEdgeNode[u] \neq null$}
      \State $\cycle \leftarrow \cycle \cup \{\inEdgeNode[u]\}$
      \State $(v,u) \leftarrow \EDGE(\inEdgeNode[u])$ 
      \State $\map[\inEdgeNode[u]] \leftarrow (v,u)$
      \State $u \leftarrow \SFIND(v)$ 
    \EndWhile
    \State Let $\sigma$ denote the weight of the maximum weight edge $(u_\sigma,v_\sigma)$ in $\cycle$.
    \State $\rep \leftarrow \SFIND(v_\sigma)$
    \For {each node $n \in \cycle$}
      \State $\reduct \leftarrow \sigma - w(\map[n])$
      \State $(u,v) \leftarrow EDGE(n)$
      \State $\SADD(v, \reduct)$
      \State $\cycleEdgeNode[\SFIND(v)] \leftarrow \cycleEdgeNode[\SFIND(v)] \cup \{n\}$
    \EndFor
    \For {each node $n \in \cycle$}
      \State $(u,v) \leftarrow EDGE(n)$
      \State $\SUNION(u, v)$
    \EndFor
    \State $\roots \leftarrow \roots \cup \{\SFIND(\rep)\}$
    \State $\varmax[\SFIND(\rep)] = \varmax[\rep]$
    \For {each node $n \in \cycle$}
      \State $(u,v) \leftarrow EDGE(n)$
      \If {$\SFIND(v) \neq \rep$}
        \State $\MELD(\queues[\rep], \queues[\SFIND(v)])$
      \EndIf
    \EndFor
  \EndIf
\end{algorithmic}
\codelinebelow

\subsubsection{Expansion phase}
We obtain the optimal arborescence from the forest $F$, which is decomposed to break the cycles of $G$.
Note that the nodes of $F$ will represent the edges of $H$ seen in Edmonds algorithm.
The expansion phase is as follows.
We first take care of the super-nodes of $F$ which are roots of the optimal arborescence, represented by the set $\rset$.
Each vertex $u$ in $\mathit{rset}$ is the representative element of a cycle, i.e. the destination of the maximum edge of a cycle.
Hence $u$ becomes a root of the optimal arborescence, and every edge incident to $u$ in $F$ must be deleted.
The tree $F$ is decomposed by deleting the node $\pi[u]$ and all its ancestors.
For the other cycles, which corresponding super-vertices are not optimal arborescence roots,
the incident edge $(u, v)$ represented by a root in $F$ is added to $H$,
and the other incident edges represented in $F$ by $\pi[v]$ and its ancestors are deleted.
The procedure ends when there are no more nodes in $F$.
The optimal arborescence is given by $H$.
The pseudo-code is as follows.

\codelineabove
\begin{algorithmic}
\State $H \leftarrow \emptyset$ \Comment {Set of edges.}
\State $R \leftarrow \{\varmax[v]\ |\ \forall v \in \rset \}$
\State $N \leftarrow$ roots of $F$
\While {$R \neq \emptyset$}
  \State $u \leftarrow \POP(R)$
  \State $N\leftarrow \texttt{DELETE-ANCESTORS}(\pi[u], N)$
\EndWhile

\While {$N \neq \emptyset$}
  \State $(u, v) \leftarrow \EDGE(\POP(N))$
  \State $H \leftarrow H \cup (u, v)$
  \State $N\leftarrow \texttt{DELETE-ANCESTORS}(\pi[v], N)$
\EndWhile   
\State \Return $H$
\end{algorithmic}

\begin{algorithmic}
\Procedure{$\texttt{DELETE-ANCESTORS}$}{$\mathit{nodeF}, N$}
  \While {$\mathit{nodeF} \neq null$}
    \For {each $\child \in \texttt{CHILDREN}(\mathit{nodeF})$}
      \State $\PARENT(\child) = \mathit{null}$
      \State $N \leftarrow N \cup \{\child\}$
    \EndFor
    \State remove $\mathit{nodeF}$
    \State $\mathit{nodeF} = \texttt{PARENT}(node)$
  \EndWhile
  \State \Return $N$
  \EndProcedure
\end{algorithmic}
\codelinebelow

\subsubsection{Illustrative example}
\begin{figure}[H]
	\centering
	\includegraphics[scale=1]{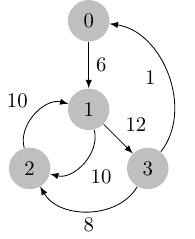}
	\caption{Input weighted directed graph.}
	\label{fig:correction_input_graph}
\end{figure}

Let us consider the graph $G=(V,E)$ in Figure~\ref{fig:correction_input_graph}.
At the beginning of the contraction phase the forest $F$ is empty.
There is a priority queue associated to each vertex and their content is $Q_0 = \{(3,0,1)\}$, $Q_1 = \{(0,1,6),(2,1,10)\}$, $Q_2 = \{(3,2,8),(1,2,10)\}$, and $Q_3 = \{(1,3,12)\}$.
We have also $\roots = \{0,1,2,3\}$, $\rset = \emptyset$, and $\varmax[v] = v$ for $v\in V$. 

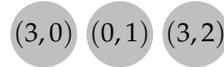
\begin{figure}[H]
	\centering
	\scalebox{1}{	
		\begin{tikzpicture}
		\node[vertex] (0) at (1,0) {$(3,0)$};
		\node[vertex] (0) at (2,0) {$(0,1)$};
		\node[vertex] (0) at (3,0) {$(3,2)$};
		\end{tikzpicture}
	}
	\caption{Forest $F$ after popping $0$, $1$ and $2$ from set $\roots$.}
	\label{fig:forest1}	
\end{figure}

We start popping vertices, denoted by $r$, from the set $\roots$ and finding the minimum weighted edge incident to each $r$.
We can safely pop $0$, $1$ and $2$ from $\roots$, and the respective minimum weight incident edges $(3,0)$, $(0,1)$, and $(3,2)$, with weights $1$, $6$ and $8$, respectively, without forming a cycle.
These edges are added to forest $F$ as nodes, leading to the state seen in Figure~\ref{fig:forest1}.
Since each vertex in $\{0,1,2\}$ forms a strongly connected component with a single vertex, we have $\pi[0]=(3,0)$, $\pi[1]=(0,1)$, and $\pi[2]=(3,2)$.

\begin{figure}[H]
	\centering
	\subcaptionbox{Cycle $\{(3,0), (0,1), (1,3)\}$ colored in green.}{
		\includegraphics[scale=1]{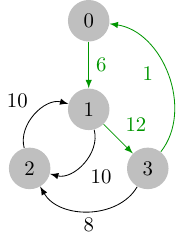}
	}\hspace{20mm}
	\subcaptionbox{Contraction result.}{
	\includegraphics{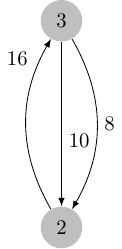}
	}
	\caption{First contraction of the input graph.}
	\label{fig:correction_contraction_1}
\end{figure}

Note that currently $roots = \{3\}$ and the content of each priority queue is $Q_0 = \emptyset$, $Q_1 = \{(2,1,10)\}$, $Q_2 = \{(1,2,10)\}$, and $Q_3 = \{(1,3,12)\}$. Vertex $3$ is then removed from set $\roots$, edge $(1,3)$ is added as a node to $F$, and $\pi[3] = (1,3)$. Also, a cycle $\{(3,0), (0,1), (1,3)\}$ is formed implying that a contraction must be performed. Let $3$ denote the cycle representant.
After the contraction we have $\varmax[3] = 3$, since $(1,3)$ is the maximum weight edge in the cycle and we have $Q_3 = \{(2,1,16)\}$, $Q_2 = \{(1,2,10)\}$, $roots = \{3\}$. Figure~\ref{fig:correction_contraction_1} depicts this first contraction.

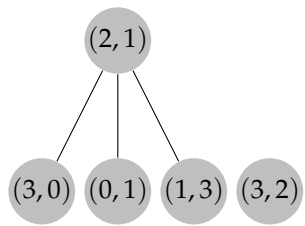
\begin{figure}[H]
	\centering
	\begin{tikzpicture}[edge/.style = {-,> = latex}]
	\node[vertex] (0) at (1,0) {$(3,0)$};
	\node[vertex] (1) at (2,0) {$(0,1)$};
	\node[vertex] (2) at (3,0) {$(1,3)$};
	\node[vertex] (3) at (4,0) {$(3,2)$};
	\node[vertex] (4) at (2,2) {$(2,1)$};
	
	\draw[edge]  (4) to (1);
	\draw[edge]  (4) to (0);
	\draw[edge]  (4) to (2);
	\end{tikzpicture}
\caption{Adding directed edges from node $(2,1)$ to the nodes of cycle $C$ in forest $F$.}
\label{fig:forest2}
\end{figure}

Vertex $3$ is yet again removed from set $roots$, moreover edge $(2,1,16)$ is popped out from $Q_3$, and added to $F$ as a node. Since edge $(2,1,16)$ is incident in a strongly connected component that contains cycle $C = \{(3,0), (0,1), (1,3)\}$, edges directed from $(2,1)$ to every edge in $C$ are created in $F$, and parent pointers are initialized in the reverse direction as shown in Figure~\ref{fig:forest2}.

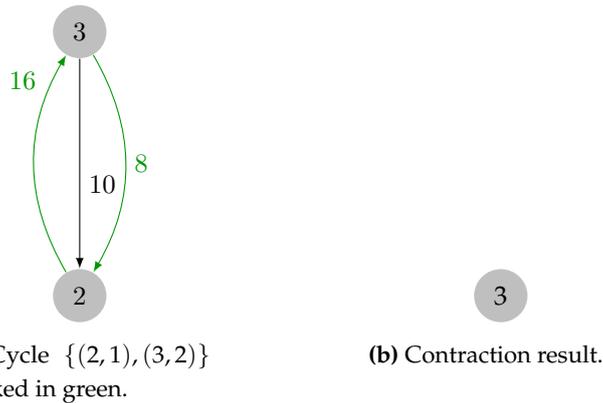
\begin{figure}[H]
	\centering
	\subcaptionbox{Cycle $\{(2,1), (3,2)\}$ marked in green.}[0.25\textwidth][c]{
		\includegraphics[scale=1]{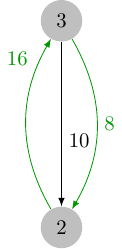}
	}\hspace{20mm}
	\subcaptionbox{Contraction result.}[0.25\textwidth][c]{
		\begin{tikzpicture}[edge/.style = {-,> = latex}]
		\node[vertex] (0) at (1,0) {$3$};
		\end{tikzpicture}
	}
	\caption{Last cycle and final contracted graph.}
	\label{fig:final_contraction}
\end{figure}

Recall that edge $(3,2)$ was previously selected and the addition of edge $(2,1)$ forms cycle $C' = \{(2,1), (3,2)\}$. After processing $C'$, let $3$ be the cycle representative, and hence $roots =\{3\}$, $Q_3 = \emptyset$ and $\varmax[3] = 3$, since $(2,1)$ is the maximum weight edge in the cycle and $\texttt{\SFIND(1)}  = 3$.
Finally $3$ is removed from set $\roots$ but $Q_3$ is empty, ending the contraction phase. The final contracted graph is presented in Figure~\ref{fig:final_contraction}.

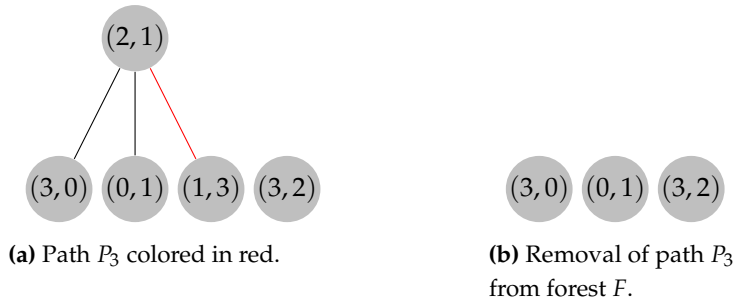
\begin{figure}[H]
	\centering
	\subcaptionbox{Path $P_3$ colored in red.} {
		\scalebox{1}{
		\begin{tikzpicture}[edge/.style = {-,> = latex}]
		\node[vertex] (0) at (1,0) {$(3,0)$};
		\node[vertex] (1) at (2,0) {$(0,1)$};
		\node[vertex] (2) at (3,0) {$(1,3)$};
		\node[vertex] (3) at (4,0) {$(3,2)$};
		\node[vertex] (4) at (2,2) {$(2,1)$};
		
		\draw[edge]  (4) to (1);
		\draw[edge]  (4) to (0);
		\draw[edge,red]  (4) to (2);
		\end{tikzpicture}
	}
	}\hspace{20mm}
	\subcaptionbox{Removal of path $P_3$ from forest $F$.}{
		\scalebox{1}{
		\begin{tikzpicture}[edge/.style = {-,> = latex}]
		\node[vertex] (0) at (1,0) {$(3,0)$};
		\node[vertex] (1) at (2,0) {$(0,1)$};
		\node[vertex] (3) at (3,0) {$(3,2)$};
		\end{tikzpicture}
		}
	}
	\caption{Forest $F$ after removing $P_3$.}
	\label{fig:forest3}
\end{figure}

The expansion phase can proceed now. Let $N = \{(2,1), (3,2)\}$, $R = \{3\}$ and $H = \emptyset$. Recall that $\pi[0]=(3,0)$, $\pi[1]=(0,1)$, $\pi[2]=(3,2)$, and $\pi[3]=(1,3)$. The expansion begins by evaluating the elements from set $R$, which contains only vertex $3$.
Since $\pi[3]=(1,3)$, and the path $P_3$ is constructed by following the child-to-parent direction until a root node is found, $P_3=\{(1,3),(2,1)\}$.
Then $P_3$ is removed from $F$ and the content of set $N$ is updated, $N = \{(3,2),(3,0),(0,1)\}$ as shown in Figure~\ref{fig:forest3}.

\begin{figure}[H]
	\centering
	\includegraphics{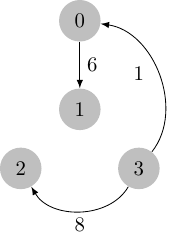}
	\caption{MSA of the input graph.}
	\label{fig:correction_arbo}
\end{figure}

Since $R = \emptyset$, the expansion phase proceeds with evaluation of nodes in set $N$.
Set $N$ is processed similarly to set $R$ with two minor changes:
the elements of $N$ when removed, are added to $H$; since $N$ contains edges $(u,v)$ as nodes, then the path $P_v$ is traced from the leaf node stored in $\pi[v]$. This process terminates when $N = \emptyset$ and $H$ holds the optimal arborescence. The final arborescence $H = \{(3,2),(3,0),(0,1)\}$ for our example is depicted in Figure~\ref{fig:correction_arbo}.

\section{Optimal dynamic arborescences}\label{sec:dynarbo}
Pollatos, Telelis and Zissimopoulos~\cite{zissimopoulosfully,pollatos2006updating} proposed two variations of an intermediary tree data structure which is built during the execution of the Edmonds algorithm on $G$
and that is then updated when $G$ changes.
We will present the data structure by Pollatos et al.~\cite{pollatos2006updating}, named \textit{augmented tree data structure} (ATree), that encodes the set of edges $H$ introduced in the previous section, along with all vertices (simple and contracted) processed during the contraction phase of Edmonds algorithm.
When $G$ is modified, the ATree is decomposed and processed, yielding a partially contracted graph $G'=(V', E')$.
Then the Edmonds algorithm is executed for $G'$. Note that only $G$ and the ATree are kept in memory.

Let us assume that the graph $G = (V, E)$ is strongly connected and that $w(u,v) > 0$, for all $(u,v) \in E$.
If $G$ is not strongly connected, we can add a vertex $v_\infty$ and $2 n$ edges such that $w(v_\infty, v) = \infty$ and $w(v, v_\infty) = \infty$, for all $v \in V$.

\subsection{ATree}
\begin{figure}[H]
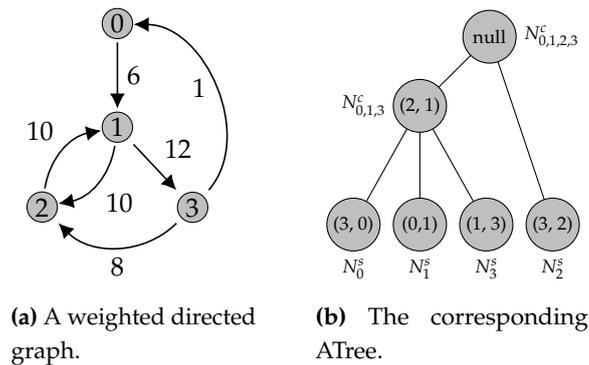

\centering
  \subcaptionbox{A weighted directed graph.}{
\begin{tikzpicture}[every node/.style={transform shape}]
\input{./figures/figure.tex}
\graphFull
\end{tikzpicture}}
\qquad
\subcaptionbox{The corresponding ATree.}{
\begin{tikzpicture}[scale=0.7, every node/.style={transform shape}]
\input{./figures/figure.tex}
\aTreeFull
\end{tikzpicture}
  }
\caption{A graph and its ATree. The root represents the edge incident to the contraction of all graph nodes ($\mathit{null}$).}
\label{figure:atree}
\end{figure}

A simple node of the ATree, represented as $N^s_v$ encodes an edge with target $v\in V$.
A complex-node, represented as $N^c_{i\ldots j}$ encodes an edge which target a super-vertex that represents a contraction of the vertices $i\ldots j\in V$.
In what follows, whenever the type of an ATree node is not known or relevant in the context, we just use $N$ to represent it.
The parent of an ATree node is the complex-node which edge targets the super-vertex into which the child edge target is contracted.
Since $G$ is strongly connected, all vertices will eventually be contracted into a single super-vertex and the ATree will have a single root.
A $\mathit{null}$ edge is encoded in the ATree root node. See Figure~\ref{figure:atree}.
The ATree takes $O(n)$ space and its construction can be embedded into the Edmonds algorithm implementation without affecting its complexity.
Let us detail how an update in $G$ affects the ATree $F$, namely edge insertions and deletions. Edge weight updates are easily achieved by deleting the edge and adding it again with the new weight. Vertex deletions are solved by deleting all related edges, and vertex insertions are trivial solved by considering $G'$ with the existing super-vertex and the new vertex (and related edges).

\subsection{Edge deletion}
Let $(u,v) \in E$ be the edge we want to delete from $G$.
If $(u,v) \notin F$, we just remove it from $G$. 
If $(u,v) \in F$, we remove $(u,v)$ from $G$ and we decompose the ATree:
we delete the node $N$ which represents the edge $(u,v)$ and, as we broke the cycles containing $(u,v)$, we also delete every ancestor node of $N$ in $F$.
Each child of a removed node becomes the root of its sub-tree.
Then, we create a partially contracted graph $G'$ with the remaining nodes in the ATree and we run the Edmonds algorithm for $G'$ to rebuild the full ATree $F$, and find the new optimal arborescence.

The graph $G' = (V', E')$ is obtained from the decomposed ATree as follows.
Note that if a complex-node $N^c_{i,...,j}$ is a root of $F$, the super-vertex representing the contraction of vertices $i,...,j$ belongs to $V'$.
Let then $\{N_{x_1},\ldots, N_{x_\ell}\}$ be the roots of the ATree $F$, where $x_k$ is the representant of the contraction when $N_{x_k}$ is a complex-node in the ATree.
Then $V' = \{x_1,\ldots,x_\ell\}$.
$E'$ is the set of the incident edges in $V'$.

\subsection{Edge insertion}
\begin{figure}[H]
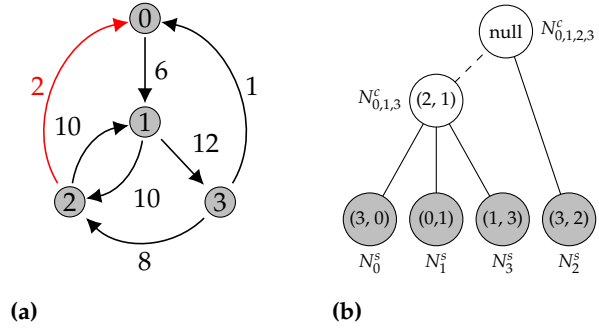

\centering
\subcaptionbox{}{
\begin{tikzpicture}[every node/.style={transform shape}]
\input{./figures/figure.tex}
\graphFullADD
\end{tikzpicture}
}
\qquad
\subcaptionbox{}{
\begin{tikzpicture}[scale=0.7, every node/.style={transform shape}]
\input{./figures/figure.tex}
\aTreeADD
\end{tikzpicture}
}
\caption{We add the edge $(2, 0)$ with weight $2$ to the graph ((a) edge represented in red).
The process starts with the analysis of node $N^s_0$, which represents edge $(3, 0)$ with weight $1$.
$(2,0)$ cannot replace $N^s_0$ edge as it is heavier.
Then $N^s_0$ parent is examined, $N^c_{0, 1, 3}$.
The corresponding edge is heavier than $(2,0)$ and $N^s_2$ is not present in its sub-tree.
Then, $(2,0)$ should replace this node, and $N^c_{0, 1, 3}$ and its ancestors are virtually deleted ((b), nodes represented in white).
The Edmonds algorithm is executed on the remaining nodes (represented in grey).}\label{figure:graphADD}
\end{figure}

Inserting a new edge $(u, v)$ is handled by reducing the problem to an edge deletion.
We first add $(u,v)$ in $G$.
Then we check if $(u,v)$ should replace an edge present in ATree $F$.
Starting from the leaf $N^s_v$ of the ATree $F$ representing an edge incident to $v$, and then following its ancestors, we check if $w(u,v)$ is smaller than the weight of the edge represented by each $N$ (see Figure~\ref{figure:graphADD}).
We can replace an edge if the previous condition holds and if $N^s_u$ is not present in its sub-tree, i.e. $(u,v)$ should not be an edge connecting two nodes of the current cycle.
We then engage a virtual deletion of the candidate node (the edge is not deleted but the ATree is decomposed), we build the graph $G' = (V', E' \cup \{(u,v)\})$ from the decomposed ATree, and we execute the Edmonds algorithm for $G'$ to rebuild the full ATree $F$.

\subsection{ATree data structure}
\label{DynImpl}
ATree is an extension of the forest $F$ data structure presented in Section~\ref{sec:tarjan}.
The nodes of the ATree maintain the following records:
the edge of $G$ the node $N$ represents, $\EDGE(N)$,
the cost of the edge at the time it was selected, $\texttt{w}_N$,
its parent $\PARENT(N)$, the list of its children $\CHILDREN(N)$,
its $\texttt{kind}$ (simple or contracted),
the list of $\texttt{contracted-edges}$ during the creation of the super-node,
the edge of maximum weight in the cycle $\texttt{e}_{\mathit{max}}$
and its weight $\texttt{w}_{\mathit{max}}$.

In an edge deletion, the edge is removed from the $\texttt{contracted-edges}$ list into which it belongs.
In the process of decomposing and reconstructing the ATree, the set of edges $E'$ corresponds to the concatenation of the lists \texttt{contracted-edges} associated to each deleted ATree nodes.
And we need to update the weight of every edge $(u, v)$ of $E'$.
Let $N^s_v$ be the simple node whose \texttt{contracted-edges} contains $(u,v)$.
The new weight $w'(u,v)$ is
$w'(u,v) = w(u,v) - \sum_{N \in P} w_N$
where $P$ is the set of ancestors of $N^s_v$,
$w(u,v)$ is the original weight, 
and $w_N$ the weight of the edge represented by $N$ at the time it was selected.
We run a BFS on each tree to find the subtracted sum $r_i$ of each simple node $N^s_i$
in $O(n)$ time. Then, we scan the edges $e$ to assign the reduced cost
$w'(e) = w(e) - r_i$.

While adding an edge $(u,v)$, we look for a candidate node to replace in the ATree. 
The process starts with $N^s_v$ and it checks every ancestor until the root is inspected or
if a node $N$ verifies $w'(\EDGE(N)) > w'(u,v)$, where $w'(\EDGE(N))$ denotes the reduced cost of the edge presented by $N$.
If the root is reached without verifying the condition,
we insert $(u,v)$ in the $\texttt{contracted-edges}$ list of
the lowest common ancestor of $N^s_{v}$ and $N^s_{u}$.
If the condition is met, we found a candidate node $N$ where $(u,v)$ could be added,
but we must determine if $(u,v)$ is safe to be added.
We check if $N^s_{u}$ is already present with a BFS in the sub-tree of root $N$.
If we find $N^s_{u}$, we insert $(u,v)$ in the $\texttt{contracted-edges}$ list of
the lowest common ancestor of $N^s_{v}$ and $N^s_{u}$.
Otherwise, we engage a virtual deletion of $\EDGE(N)$
(the edge is not deleted but the ATree is decomposed),
then we build the graph $G'=(V', E'\cup \{(u,v)\})$ and execute the Edmonds algorithm for $G'$ as mentioned before.
The pseudo-code for finding a candidate is as follows, where $(u,v)$ is edge to be inserted.
\begin{algorithmic}
\State $\nodeF = N^s_{v}$
\If {$w(\EDGE(\nodeF)) > w (u,v)$}
  \State \Return $\nodeF$
\EndIf
\State $S \leftarrow \emptyset$ \Comment{Let $S$ be a set.}
\While {$\nodeF \neq \mathit{null}$}
  \State $S = S\cup \{ \nodeF\})$
  \State $\nodeF = \PARENT(\nodeF)$
\EndWhile
\State Let $L$ be a LIFO containing the nodes in cycle creation order.
\State $\mathit{compare} = \mathit{false}$
\State $\mathit{candidate} = \mathit{null}$
\While {$L\neq \emptyset$}
  \State $\nodeF = \texttt{POP}(L)$
  \If {$\nodeF$ is root}
    \State \Return $\mathit{null}$
  \EndIf
  \If {$\nodeF\in S$}
    \State $\mathit{compare} = \mathit{true}$
  \EndIf
  \State $\mathit{candidate} = \texttt{FIND-CANDIDATE}(\nodeF, (u,v), \mathit{compare})$
  \State $\mathit{compare} = \mathit{false}$
  \If {$\mathit{candidate} \neq \mathit{null}$}
    \State \BREAK
  \EndIf
\EndWhile
\State \Return $\mathit{candidate}$
\end{algorithmic}

\begin{algorithmic}
\Procedure{$\texttt{FIND-CANDIDATE}$}{$\nodeF, e_{in}, \mathit{compare}$}
\State Let $w_{\varmax}$ be the maximum weight edge in $\CHILDREN(\nodeF)$.
\For {each $\mathit{child} \in \CHILDREN(\nodeF)$}
  \State $(u'', v'') = \EDGE(child)$
  \State $\mathit{cost} = w_{\varmax} - (w(u''.v'') + \texttt{SFIND-WEIGHT}(v''))$
  \State $\texttt{SADD-WEIGHT}(v'', \mathit{cost})$
\EndFor
\State $(u',v') = \texttt{EDGE}(\nodeF)$
\State $w'(u',v') = w(u',v') + \texttt{SFIND-WEIGHT}(v')$
\State $w'(u,v) = w(u,v) + \texttt{SFIND-WEIGHT}(v)$
\If {$\compare$ and $w'(u,v) < w'(u',v')$}
  \State \Return $\nodeF$
\EndIf
\For {each $\child \in \CHILDREN(\nodeF)$}
  \State $(u'', v'') = \EDGE(\child)$
  \State $\SUNION(u'', v'')$
\EndFor
\State \Return $\mathit{null}$
\EndProcedure
\end{algorithmic}

\section{Implementation details and analysis}
\label{sec:impl}
Let us detail and discuss our implementation, namely used data structures and their customization, for finding an optimal arborescence and to dynamically maintain it.
It follows the pseudo-code described in the previous sections.
As mentioned earlier, this implementation is built on the theoretical results introduced by Edmonds~\cite{edmonds1967optimum}, Tarjan~\cite{tarjan1977finding}, and Camerini et al.~\cite{camerini1979note} for the static algorithm, and on the results by 
Pollatos, Telelis and Zissimopoulos~\cite{zissimopoulosfully,pollatos2006updating} for the dynamic algorithm.
The implementation incorporates all these results, namely the contraction and expansion phases by Edmonds, the bookkeeping mechanisms proposed by Tarjan, and the forest data structure introduced by Camerini et al., and further extended as the ATree data structure.
Recall that the bookkeeping mechanism adjusted to maintain the forest data structure relies on the following data structures: 
for every node $v$, a list $L(v)$ stores each edge incident to $v$;
disjoint sets keep track of the strongly and weakly connected components;
a collection of $\queues$ keeps track of the edges entering each vertex;
and a forest or, in the dynamic case, an ATree $F$.

\subsection{Incidence lists}
Since edges of $G$ are processed by incidence and not by origin, $G$ is represented as an array of edges sorted with respect to target vertices.
This is beneficial since it takes advantage of memory locality bringing improvements to the overall performance.

\subsection{Disjoint sets}
Two implementations of the union-find data structure for managing disjoint sets are used, with both supporting the standard operations.
One is used to represent weakly connected components, while the other is employed for strongly connected components.
The latter is an augmented version.
In the case of the first implementation, the following common operations are supported: $\texttt{WFIND}(x)$ that returns a pointer to the representative element of the unique set containing $x$; $\texttt{WUNION}(x)$ which unites the sets that contain $x$ and $y$; and $\texttt{WMAKE-SET}(x)$ that creates a new set whose only element and representative is $x$.
For the augmented implementation, the same operations are supported, but named $\texttt{SFIND}$, $\texttt{SUNION}$ and $\texttt{SMAKE-SET}$; two extra operations are also supported, namely $\texttt{SADD-WEIGHT}$ and $\texttt{SFIND-WEIGHT}$ detailed below.

Our implementations of union-find data structure rely on the conventional heuristics, namely union by rank and path compression, achieving nearly constant time per operation in practice; $m$ operations over $n$ elements take $O(m \alpha(n))$ amortized time, where $\alpha$ is the inverse of the Ackermann~\cite{tarjan1984worst}.
Both implementations use two arrays of integers, namely the rank and the parent array instead of pointer based trees.
Even though operations computational complexity is theoretically speaking the same, using arrays instead of pointers promotes again memory locality since arrays are allocated in contiguous memory.

The purpose of having a different implementation for strongly connected components is to bring a constant time solution for the computation of the reduced costs, exploiting the path compression and union by rank heuristics.
While finding the minimum weight incident edges in every vertex in the contraction phase, cycles may arise.
Then the maximum weight edge in the cycle is found, the reduced costs are computed, and the weight of incident edges is updated by summing the reduced costs.
In this context, the augmented version of the union-find data structure supports then the following operations as mentioned above: $\texttt{SADD-WEIGHT}(x, k)$ which adds a constant $k$ to the weight of all elements of the set containing $x$, and $\texttt{SFIND-WEIGHT}(x)$ that returns the accumulated $\mathit{weight}$ for the set containing $x$.
Supporting these operations requires an additional attribute $\mathit{weight}$, represented internally as a third array to store the weights.
The weights are initialized with $0$.
The $\texttt{SADD-WEIGHT}(x, k)$ operation adds value $k$ to the root or representative element of the set containing $x$ in constant time.
The operation $\texttt{SFIND}(x)$ has been rewritten for updating the weights whenever the underlying union-find tree structure changes due to the path compression heuristic; this change does not change the complexity of this operation.
The operation $\texttt{SFIND-WEIGHT}(x)$ performs the sum of all values stored in field $\mathit{weight}$ on the path from $x$ until we meet the root of disjoint-set containing $x$; the cost of this operation is identical to the cost of operation $\texttt{SFIND}$.
A constant time solution is obtained then for updating the weight of all elements in a given set, which allows us to update the weight of all edges incident on a given vertex also in constant time.

\subsection{Queues}
Heaps are used to implement the priority queues which track the edges incident in each vertex. 
In this context, three types of heaps were implemented and tested, namely binary heaps~\cite{williamsalgorithm},
binomial heaps~\cite{vuillemin1978data} and pairing heaps~\cite{fredman1986pairing,pettie2005towards}.
The pairing heaps is the alternative that has simultaneously better theoretical and expected experimental results; although binary heaps are faster than all other heap implementations when the decrease-key operation is not needed, pairing heaps are often faster than $d$-ary heaps (like binary heaps) and almost always faster than other pointer-based heaps~\cite{larkin2014back}.
Our experimental results consider also this comparison (see Section~\ref{sec:res}).
With respect to theoretical results, using pairing heaps to implement priority queues, and assuming that $n$ is the size of a heap, the common heap operations are as follows: $\texttt{INIT}(L)$ creates a heap with elements in list $L$ in $O(n)$ time; $\texttt{INSERT}(h,e)$ inserts an element $e$ in the heap $h$ in $\Theta(1)$ time; $\texttt{GET-MIN}(h)$ obtains the element with minimum weight in $\Theta(1)$ time; $\texttt{EXTRACT-MIN}(h)$ returns and removes from the heap $h$ the element with minimum weight in $O(\log n)$ amortized time; $\texttt{DECREASE-KEY}(h,e)$ decreases the weight of element $e$ in $o(\log n)$ amortized time; and $\texttt{MELD}(h_1, h_2)$ merges two heaps $h_1$ and $h_2$ in $\Theta(1)$ time. 

Our implementation does not rely on the $\texttt{DECREASE-KEY}$ operation, but it relies heavily on the $\texttt{MELD}$ and 
$\texttt{EXTRACT-MIN}$ operations.
In this context it is important to note that the $\texttt{MELD}$ operation takes $O(n)$ time for binary heaps and $O(\log n)$ time for binomial heaps.
The $\texttt{EXTRACT-MIN}(h)$ runs in $O(\log n)$ time for both binary and binomial heaps.
On the other hand both pairing and binomial heaps are pointer based data structures, while binary heaps are array based.
Hence, it is not clear a priori which heap implementation would be better in practice and, hence, it is a topic of analysis in our experimental evaluation as mentioned.

\subsection{Forest}
Several data structures were introduced to manage $F$ and the cycles in $G$.
A set $\rset$ holds the roots of the optimal arborescence, i.e. the vertices which do not have any incident edge.
A table $\varmax$ holds the destination of the maximum edges in a strongly connected component.
A table $\pi$ points to the leaves of $F$, where $\pi[v] = (u, v)$ means that the node $(u, v)$ of $F$ was created during the
evaluation of vertex $v$.
The table $\mathit{inEdgeNode}$ holds for each $v$, the unique node of $F$ entering the strongly connected component represented by $v$.
Finally, the list $\mathit{cycleEdgeNode}$ holds the lists of nodes in a cycle, where $\mathit{cycleEdgeNode}[rep]$ holds the nodes of the cycle represented by $rep$.

These data structures allow us to construct and maintain the forest $F$ within the contraction phase without burdening the overall complexity of the algorithm. They allow also to extract an optimal arborescence in linear time during the expansion phase. Detailed pseudo-code has been presented in Section~\ref{sec:tarjan}.

This representation is extended for implementing the ATree taking into account the data structure description and the pseudo-code presented in Section~\ref{DynImpl}.

\subsection{Complexity}
Let us discuss the complexity of our implementation for finding a (static) optimal arborescence in a graph $G$ with $n$ vertices and $m$ edges.

In the initialization phase we mainly have the $n$ $\texttt{INIT}$ operations for the priority queues, the $n$ $\texttt{SMAKE-SET}$ operations on augmented disjoint sets, the $n$ $\texttt{WMAKE-SET}$ operations on disjoint sets, and $O(n)$ operations on other data structures. All these operations take constant time each, thus the initialization takes $O(n)$ time.

In the contraction phase, only the operations on priority queues and disjoint sets may not take constant time.
The operations on priority queues are at most $m$ $\texttt{EXTRACT-MIN}$ operations and $n$ $\texttt{MELD}$ operations.
Since $\texttt{EXTRACT-MIN}$ takes $O(\log n)$ time and (for pairing heaps) the $\texttt{MELD}$ operation takes constant time, then it takes $O(m\log n)$ total time for maintaining priority queues.
The operations on disjoint-sets are $m$ $\texttt{WFIND}$ and $\texttt{SFIND}$ operations,
$n$ $\texttt{WUNION}$ and $\texttt{SUNION}$ operations, and $n$ $\texttt{SADD-WEIGHT}$ operations.
The disjoint set operations take $O(m \alpha(n))$ total time where $\alpha$ is the inverse of the Ackermann function~\cite{tarjan1984worst}.
The other operations run in $O(m + n)$ time.
Therefore, the contraction phase takes $O(m\log n)$ time. 

In the expansion phase, $F$ contains no more than $2n -2$ nodes and each node of $F$ is visited exactly once,
so the procedure takes $O(n)$ time.
The total time required to find an optimal arborescence is therefore dominated by the priority queue operations yielding a final time complexity of $O(m \log n)$.

Let us analyse now the cost of maintaining dynamically the optimal arborescence.
Let $\rho$ be the set of affected vertices and edges, $|\rho|$ the number of affected vertices, and $|| \rho ||$ the number of affected edges.
A vertex is affected if it is included in a different contraction in the new output after an edge insertion or removal.
Note that $|\rho| < n$, that all operations in an addition or deletion of an edge
occur in $O(n)$ time and that a re-execution of Edmonds algorithm processes only the affected vertices.
The update of an optimal arborescence, using the implementation presented in Section~\ref{sec:tarjan},
can then be achieved in $O(n + || \rho || \log |\rho|)$ time per edge insertion or removal.

\section{Experimental evaluation}
\label{sec:res}
We implemented the original Edmonds algorithm as described in Section~\ref{algo:edmonds}, and Tarjan algorithm as described in Section~\ref{sec:tarjan}.
The implementation of Tarjan algorithm has three variants which differ only on the heap implementation.
As discussed before, we considered binary heaps, binomial heaps and pairing heaps in our experiments.
Algorithms were implemented in Java $11$, and binaries were compiled with \texttt{javac} $11.0.20$.
Experiments were performed on a computer with the following hardware:
Intel(R) Xeon(R) Silver 4214 CPU @ 2.20GHz and $16$ GB of RAM.

The aim of this experimental evaluation is to compare the performance of Edmonds original algorithm with Tarjan algorithm, to evaluate the use of different heap implementations, and to investigate the practicality of the dynamic algorithm for dense and sparse graphs.
As datasets we used randomly generated graphs, both dense and sparse, and real phylogenetic data.

\subsection{Datasets}
Graphs datasets are comprised  by sparse and dense graphs generated accordingly to well known random models.
To generate {\em sparse} graphs we considered three different models.
One of them was the Erdos-Rényi (ER) model~\cite{gilbert1959random}, with $p = \frac{c \log n}{n}$, $c \geq 1$, where $p$ denotes the probability of linking a node $u$ with a node $v$ and $n$ is the number of nodes in the network.
Whenever $p$ has the previously defined value, the network has one giant component and some isolated nodes.
Moreover, these graphs were generated  using \verb+fast_gnp_random_graph+ generator of the NetworkX library~\cite{SciPyProceedings_11}, with $p=0.02$.

Sparse scale-free directed graphs were also generated using the model by Bollob{\'a}s et al.~\cite{bollobas2003directed} (identified as {\em scale-free} in our experiments) and
a variant of the {\em duplication} model by Chung et al.~\cite{DBLP:journals/jcb/ChungLDG03}.
The first were generated using the \verb+scale_free_graph+ function of the NetworkX library, with all parameters set with their omission value except the number of nodes.
The later were generated using our own implementation, where given $0\leq p \leq 1$, the partial duplication model
builds a graph $G = (V, E)$ by partial duplication as follows: start with
a single vertex at time $t = 1$ and, at time $t > 1$, perform a duplication step:
uniformly select a random vertex $u$ of $G$;
add a new vertex $v$ and edges $(u, v)$ and $(v, u)$ with (different) random weights;
for each neighbor $w$ of $u$, add edges $(v, w)$ and/or $(w, v)$ with probability $p$, and random integer weights chosen uniformly from $[0, 1000]$.

Dense graphs were generated using the \verb+complete_graph+ generator of the NetworkX library, that creates a {\em complete} graph, i.e. all pairs of distinct nodes have an edge connecting them. Edge weights were assigned randomly.

Running time and memory is averaged over five runs and for five different graphs of each size, for all models.

We used also real phylogenetic data in the dynamic updating evaluation, namely real dense graphs using phylogenetic datasets available on EnteroBase~\cite{zhou2020enterobase}, respective details are shown in Table~\ref{table:dataset}.
Graphs were built based on the pairwise distance among genetic profiles, as usual in distance based phylogenetic inference~\cite{vaz2021distance}.
The experiments on these data were carried out by considering increasing volumes of data, namely $[10\%, 20\%, 30\%, ..., 100\%]$.
\begin{table}[H]
\caption{Phylogenetic datasets. The first three without missing data. The number of vertices $n$ is the number of genetic profiles in each dataset.}
\label{table:dataset}
\centering
\begin{tabular}{lrr}
   \toprule
Datasets                   & $n=|V|$           & $m=|E|$  \\ \midrule
\textit{clostridium.Griffiths}        & $440$            & $193600$     \\
\textit{Moraxella.Achtman7GeneMLST}  & $773$            & $597529$     \\
\textit{Salmonella.Achtman7GeneMLST} & $5464$           & $29855296$   \\
\textit{Yersinia.McNally}             & $369$      & $136161$  \\
\bottomrule
\end{tabular}
\end{table}
\subsection{Edmonds versus Tarjan}
\begin{figure}[H]
	\centering
  \subcaptionbox{Running time for complete graphs.\label{fig:edmonds_eff_in_complete}}{\includegraphics[width=0.49\textwidth]{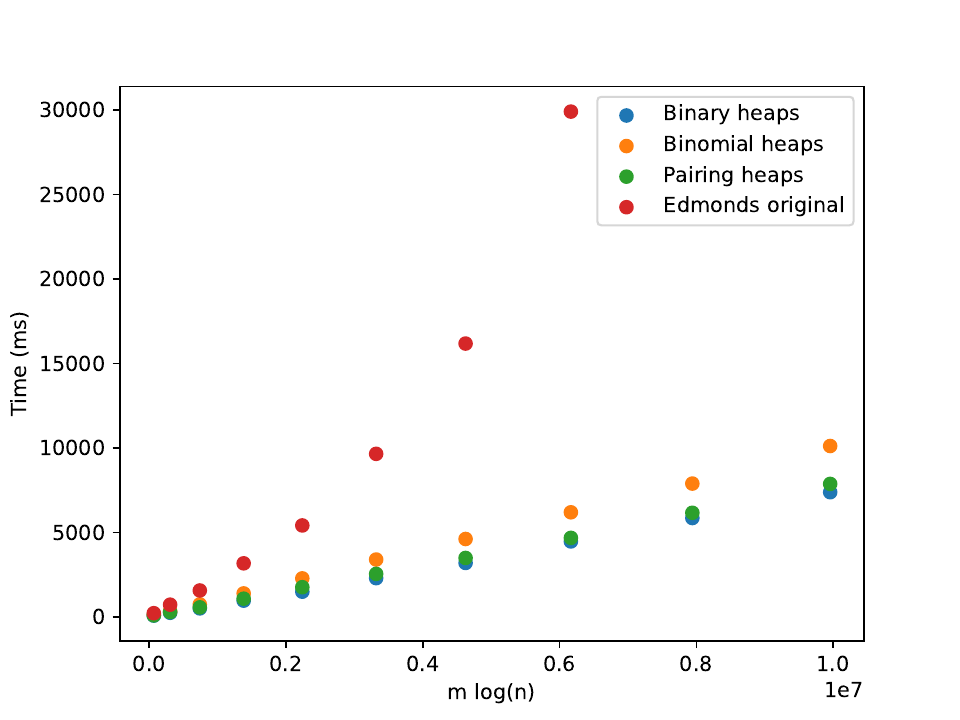}}
  \subcaptionbox{Running time for sparse graphs.\label{fig:edmonds_eff_in_sparse}}{\includegraphics[width=0.49\textwidth]{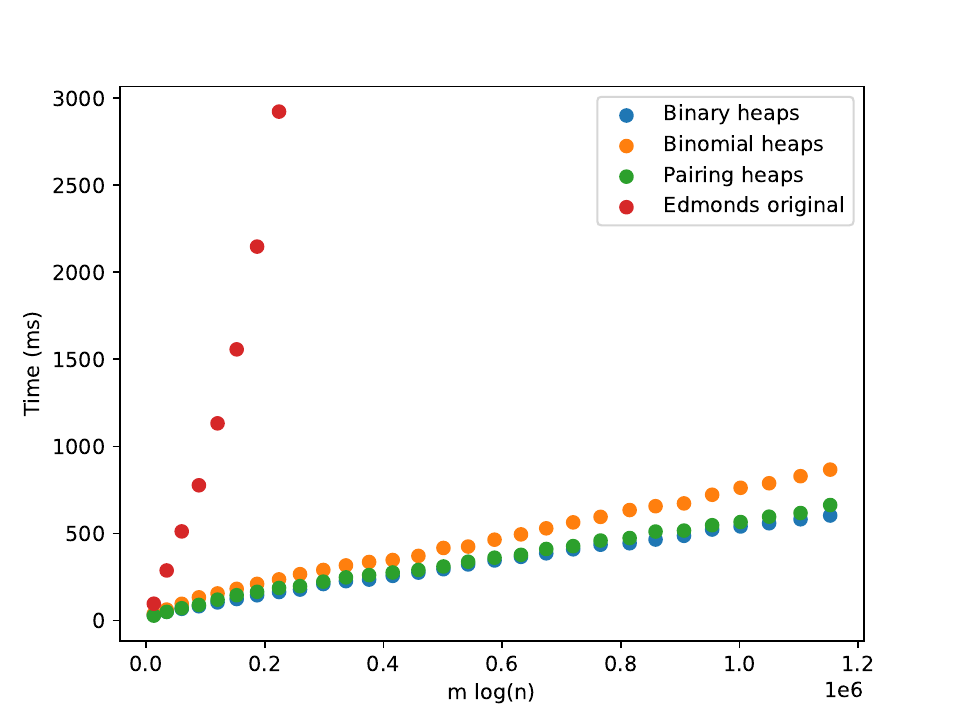}}\\[1ex]
  \subcaptionbox{Memory for complete graphs.\label{fig:edmonds_eff_in_complete_mem}}{\includegraphics[width=0.49\textwidth]{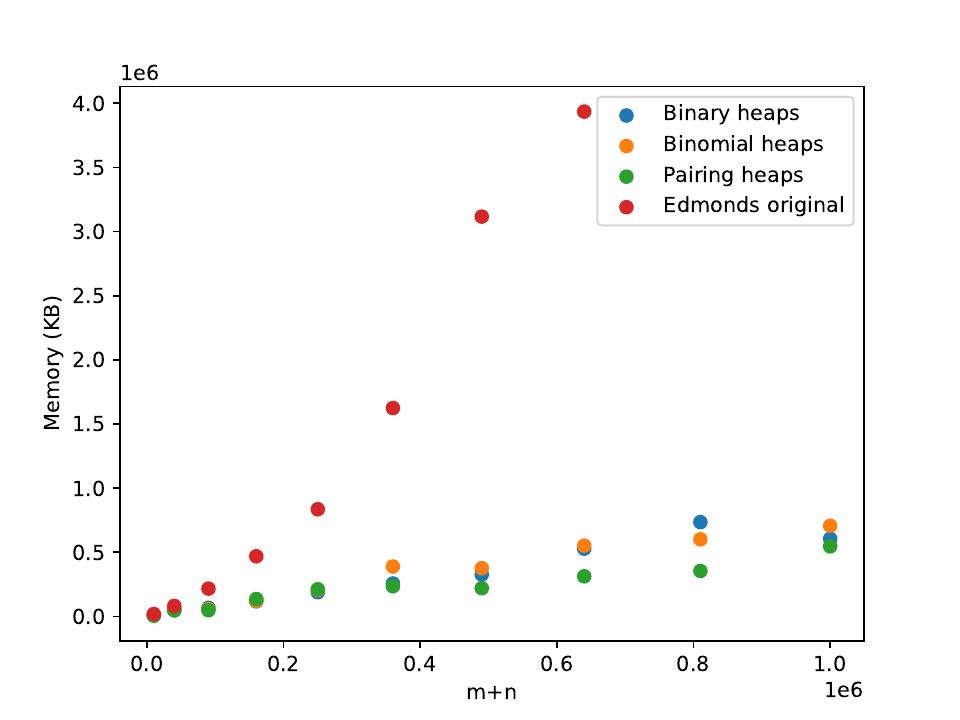}}
  \subcaptionbox{Memory for sparse graphs.\label{fig:edmonds_eff_in_sparse_mem}}{\includegraphics[width=0.49\textwidth]{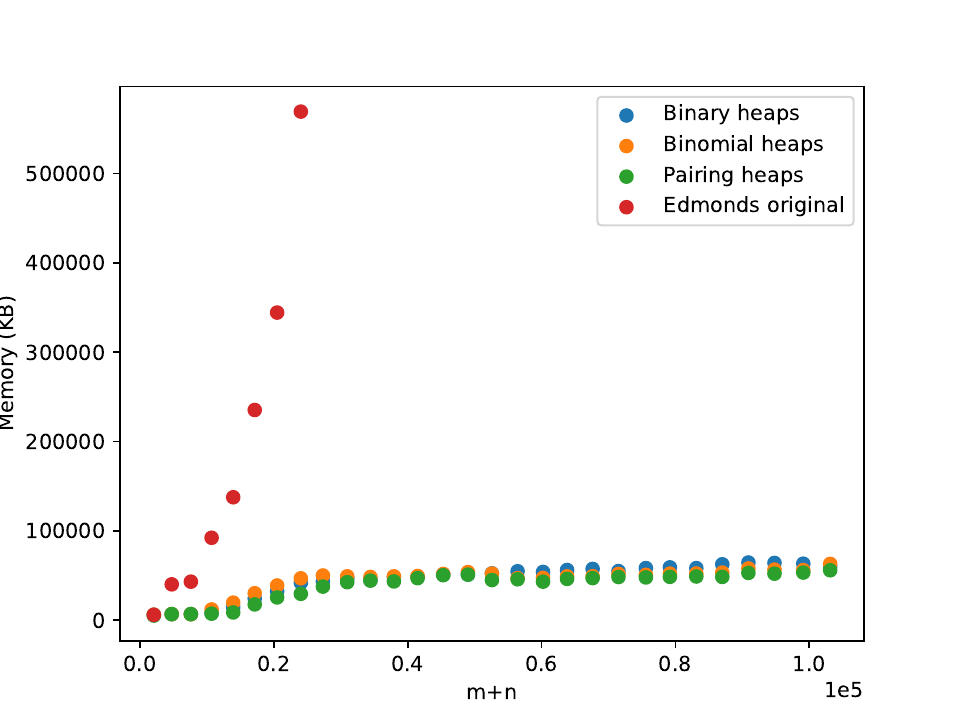}}
	\caption{Comparison between the Edmonds algorithm and the Tarjan algorithm (three different heap implementation) on complete and sparse graphs.}
	\label{fig:eff_vs_in}
\end{figure}

We compared both Edmonds and Tarjan algorithms for complete and sparse graphs using generated graph datasets.
This comparison is presented in~Figure \ref{fig:eff_vs_in}. As expected, Tarjan algorithm is faster and the experimental running time follows  the expected theoretical bound of $O(m\log n)$.
The memory requirements are also lower for the Tarjan algorithm, growing linearly with the size of the graph, as expected.

Given these results, we omit Edmonds algorithm from the remaining evaluation.

\subsection{Different heap implementations}
\begin{figure}[H]
	\centering
  \subcaptionbox{Running time for scale-free graphs.\label{fig:edmonds_eff_in_scale_free}}{\includegraphics[width=0.49\textwidth]{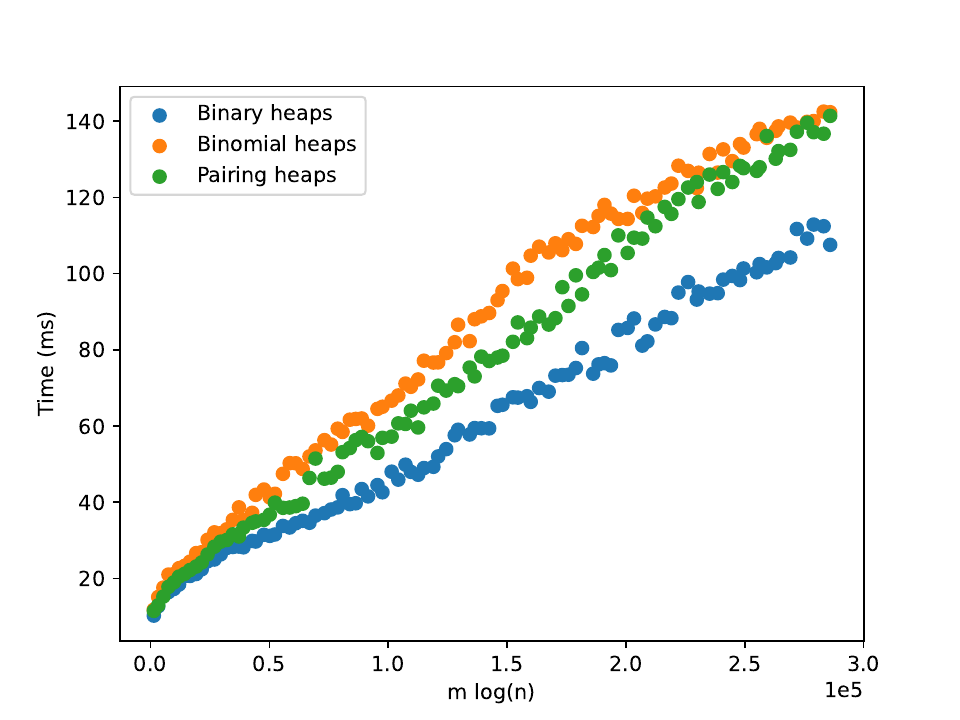}}
  \subcaptionbox{Running time for duplication model graphs.\label{fig:edmonds_eff_in_duplication}}{\includegraphics[width=0.49\textwidth]{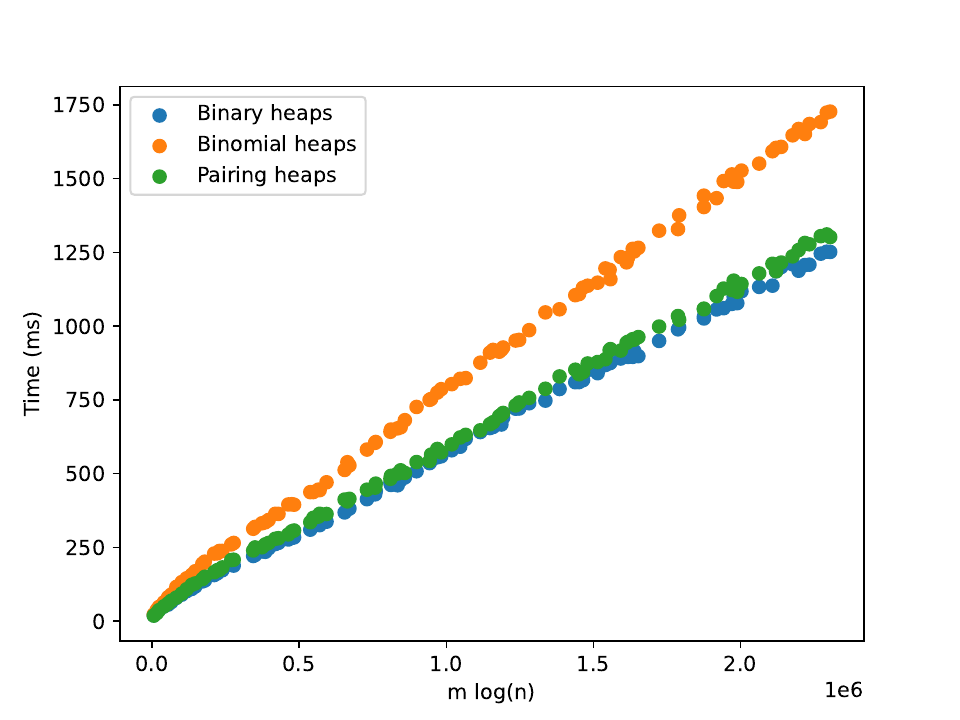}}\\[1ex]
  \subcaptionbox{Memory for scale-free graphs.\label{fig:edmonds_eff_in_scale_free__mem}}{\includegraphics[width=0.49\textwidth]{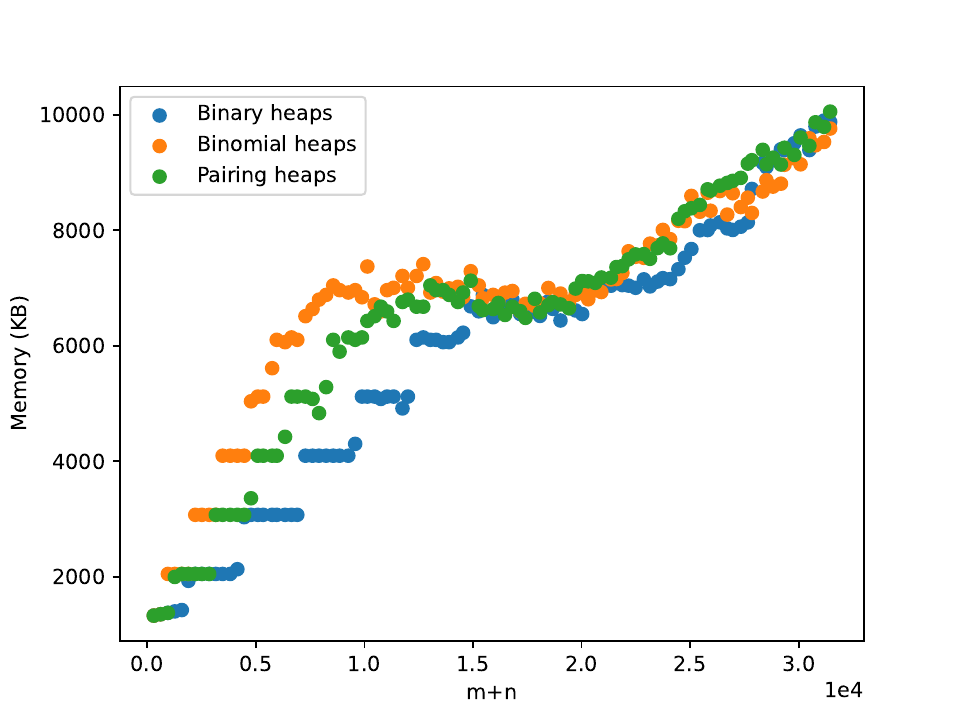}}
  \subcaptionbox{Memory for duplication model graphs.\label{fig:edmonds_eff_duplication_mem}}{\includegraphics[width=0.49\textwidth]{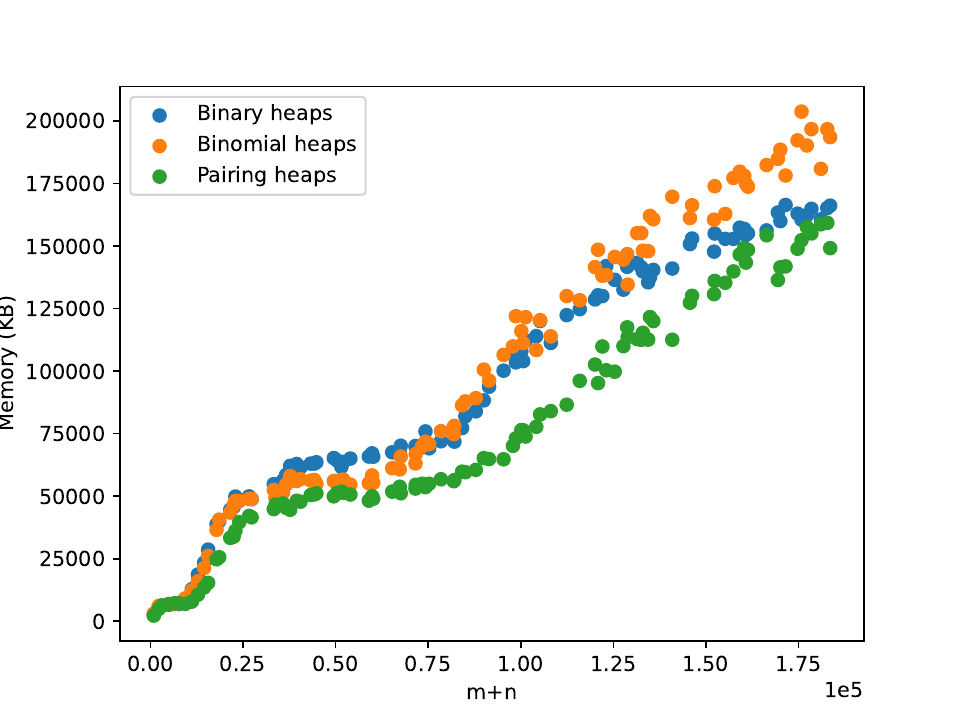}}
	\caption{Comparison if three different heap implementations within the Tarjan algorithm sparse scale-free graphs.}
	\label{fig:eff_vs_in_2}
\end{figure}

Results for scale-free graphs are presented in Figure~\ref{fig:eff_vs_in_2}.
The running time and memory requirements are according to expectations and to the analysis for complete and sparse graphs in the previous section.
The somewhat strange behavior in memory plots for a lower number of vertices and edges is due to Java's garbage collector and it can be ignored.

The focus in this section is the performance of different heap implementations together with Tarjan algorithm. The improved theoretical performance of binomial and pairing heaps is not supported by our experiments and in fact fared no better than binary heaps. Pairing heaps obtained a similar time performance to binary heaps in the duplication models while simultaneously using less space. This is particularly interesting since the meld operation is more efficient for pairing heaps. However the
memory locality exploited by binary heaps plays here an important role.

\subsection{Dynamic optimal arborescences}
Let us compare the performance of maintaining dynamic optimal arborescences versus ab initio computation on edge updates.
Both implementations rely on the algorithm by Tarjan described in this paper.
Our experiments consist on evaluating the running time and required memory for adding and deleting edges.
Results are averaged over a sequence of $10$ independent \texttt{DELETE} operations, and also over a sequence of $10$ independent \texttt{ADD} operations.
The sequences of edges subject to deletion or insertion were randomly selected.

\begin{figure}[H]
	\subcaptionbox{\texttt{DELETE} operations.\label{fig:dynamic_remove}}{\includegraphics[scale=0.46]{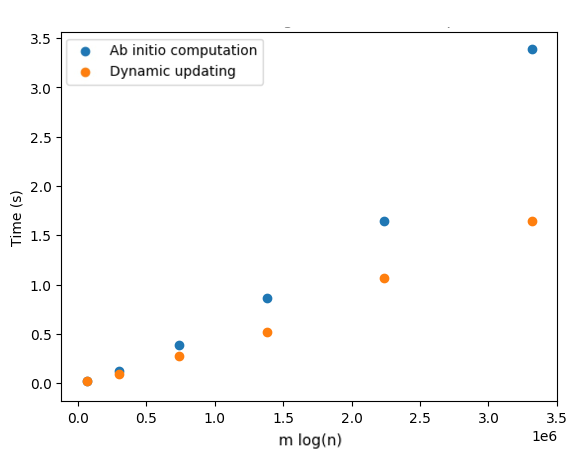}
	}
	\subcaptionbox{\texttt{ADD} operations.\label{fig:dynamic_add}}{\includegraphics[scale=0.46]{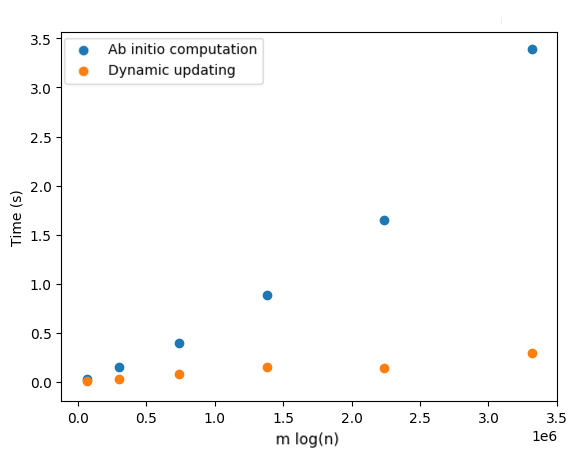}
	}
  \caption{Optimal arborescence updating versus ab initio computation for \texttt{DELETE} and \texttt{ADD} operations on complete graphs. Running time averaged over $10$ random operations.}
\end{figure}

\begin{figure}[H]
	\subcaptionbox{\texttt{DELETE} operations.\label{fig:dynamic_delete_clostridium}}{\includegraphics[scale=0.46]{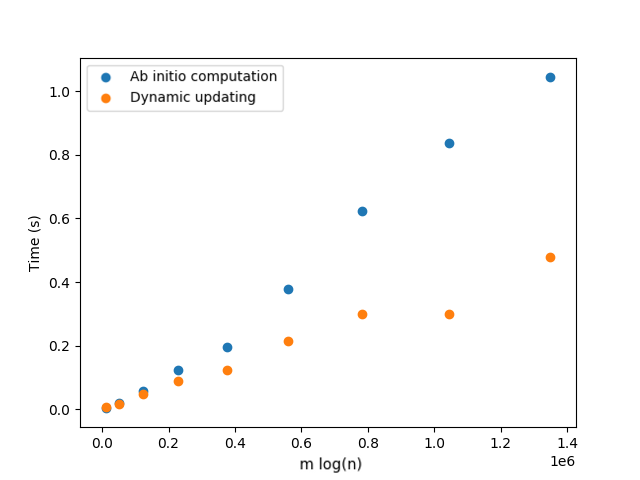}
	}
	\subcaptionbox{\texttt{ADD} operations.\label{fig:dynamic_add_clostridium}}{\includegraphics[scale=0.46]{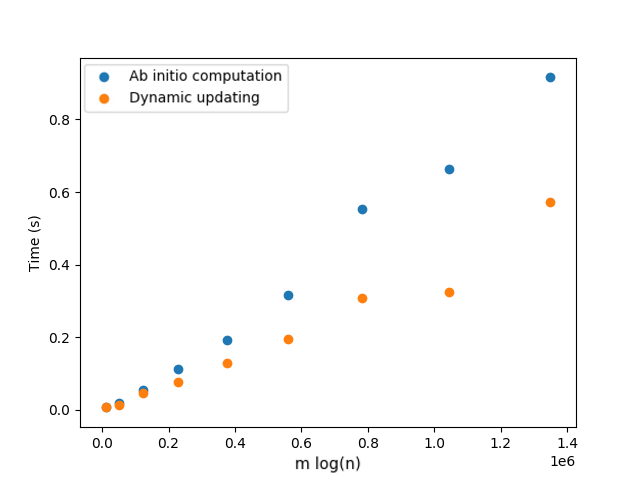}
	}
	\caption{Optimal arborescence updating versus ab initio computation for \texttt{DELETE} and \texttt{ADD} operations on \textit{clostridium.Griffiths} dataset. Running time averaged over $10$ random operations.}
\end{figure}

\begin{figure}[H]
	\subcaptionbox{\texttt{DELETE} operations.\label{fig:dynamic_delete_moraxella}}{\includegraphics[scale=0.46]{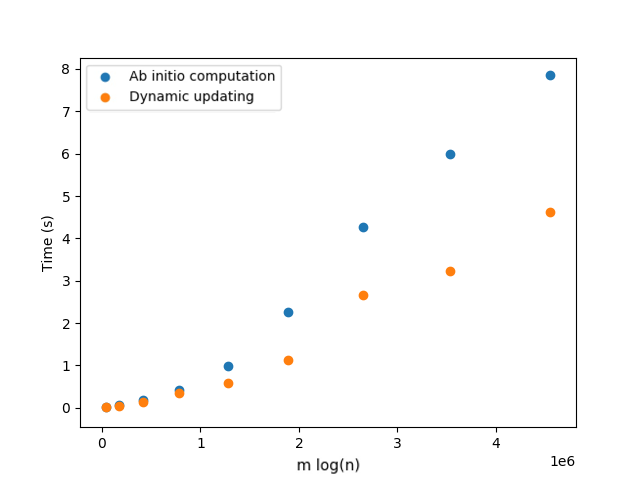}
	}
	\subcaptionbox{\texttt{ADD} operations.\label{fig:dynamic_add_moraxella}}{	\includegraphics[scale=0.46]{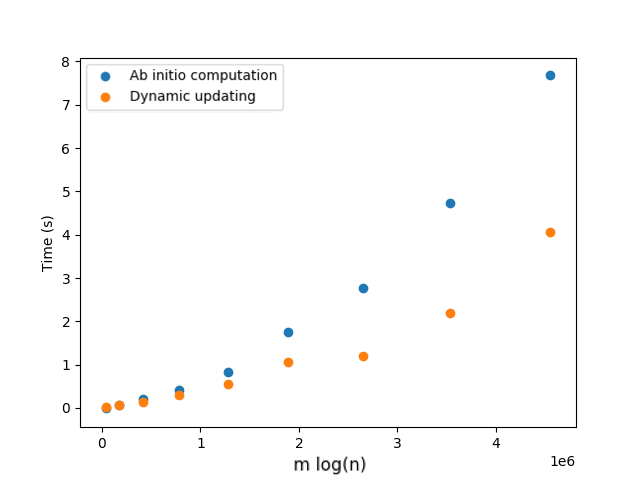}
	}
	\caption{Optimal arborescence updating versus ab initio computation for \texttt{DELETE} and \texttt{ADD} operations on \textit{Moraxella.Achtman7GeneMLST} dataset. Running time averaged over $10$ random operations.}
\end{figure}

\begin{figure}[H]
		\subcaptionbox{\texttt{DELETE} operations.\label{fig:dynamic_delete_salmonella}}{	\includegraphics[scale=0.46]{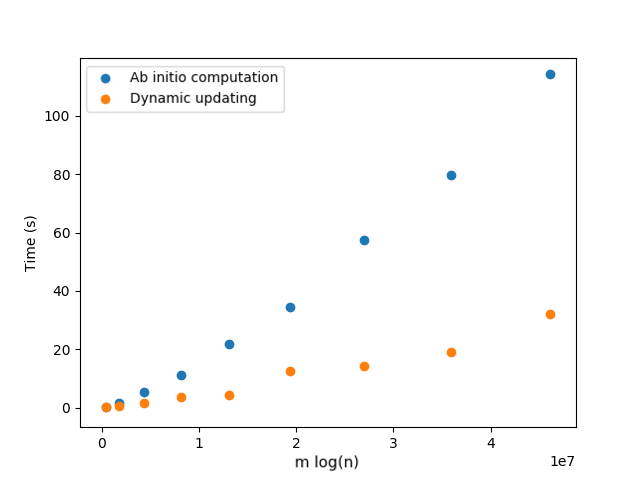}
	}
	\subcaptionbox{\texttt{ADD} operations.\label{fig:dynamic_add_salmonella}}{	\includegraphics[scale=0.46]{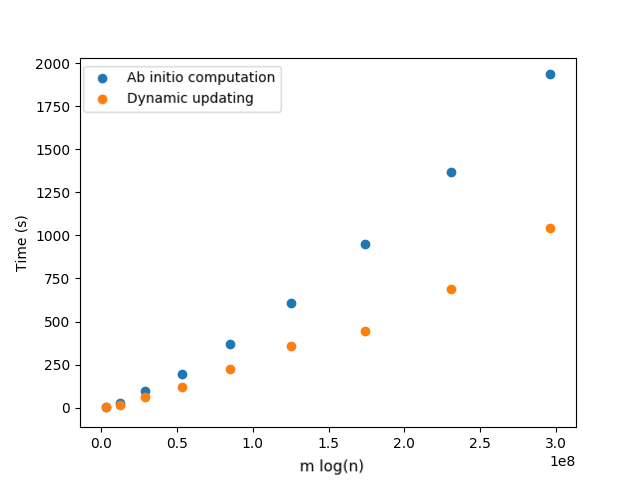}
	}
	\caption{Optimal arborescence updating versus ab initio computation for \texttt{DELETE} and \texttt{ADD} operations  on \textit{Salmonella.Achtman7GeneMLST} dataset. Running time averaged over $10$ random operations.}
\end{figure}

\begin{figure}[H]
	\subcaptionbox{\texttt{DELETE} operations.\label{fig:dynamic_delete_yersiniacg}}{\includegraphics[scale=0.46]{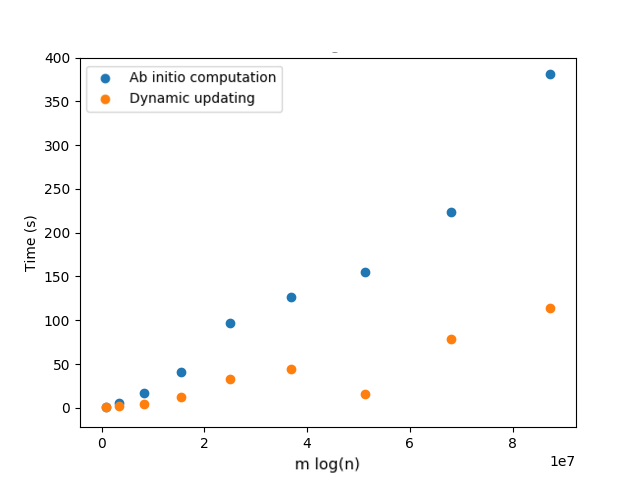}
	}
	\subcaptionbox{\texttt{ADD} operations.\label{fig:dynamic_add_yersiniacg}}{\includegraphics[scale=0.46]{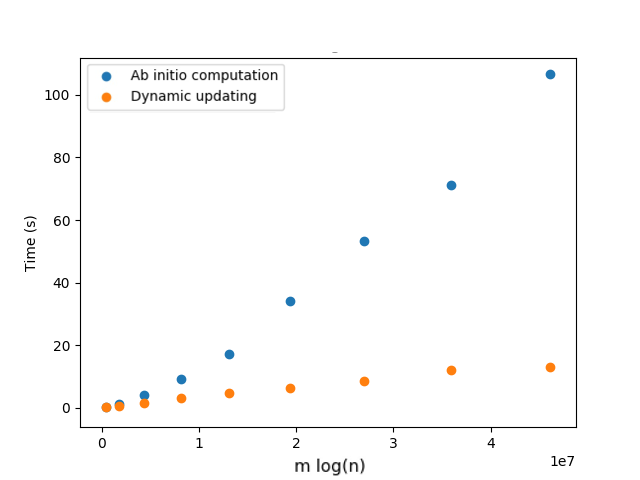}
	}
	\caption{Optimal arborescence updating versus ab initio computation for \texttt{DELETE} and \texttt{ADD} operations on \textit{Yersinia.cgMLSTv1} dataset. Running time averaged over $10$ random operations.}
\end{figure}

\begin{figure}[H]
	\subcaptionbox{\texttt{DELETE} operations.\label{fig:dynamic_delete_yersiniawg}}{\includegraphics[scale=0.46]{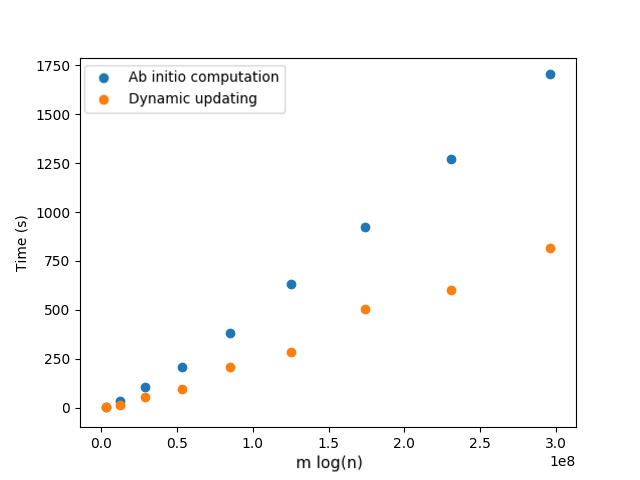}
	}
	\subcaptionbox{\texttt{ADD} operations.\label{fig:dynamic_add_yersiniawg}}{\includegraphics[scale=0.46]{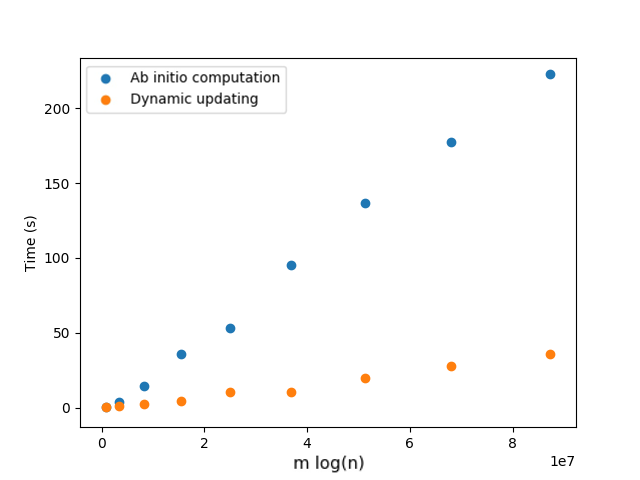}
	}
	\caption{Optimal arborescence updating versus ab initio computation for \texttt{DELETE} and \texttt{ADD} operations on \textit{Yersinia.wgMLST} dataset. Running time averaged over $10$ random operations.}
\end{figure}

Figure~\ref{fig:dynamic_remove} provides the results for the \texttt{DELETE} operations.
We observe that updating the arborescence is twice as fast compared with its ab initio computation.
Note that these results are aligned with the results presented by Pollatos, Telelis and Zissimopoulos~\cite{zissimopoulosfully,pollatos2006updating}.
Figures~\ref{fig:dynamic_delete_clostridium} to~\ref{fig:dynamic_delete_yersiniawg} provide the results for the \texttt{DELETE} operation over phylogenetic data described above.
As the size of the dataset grows, and the inferred graph becomes larger, the dynamic updating becomes also more competitive, being twice as fast when compared with the ab initio computation.

The results for the \texttt{ADD} operations are presented in Figure~\ref{fig:dynamic_add} for complete graphs, and in Figures~\ref{fig:dynamic_add_clostridium} to \ref{fig:dynamic_add_yersiniawg}.
It is clearly perceived that the ab initio computation is outmatched by the dynamic updating, in particular as the size of the graph grows.
The dynamic updating is consistently at least twice as fast as the ab initio computation, surpassing often that speedup factor.

\begin{figure}[H]
	\centering
		\includegraphics[scale=0.5]{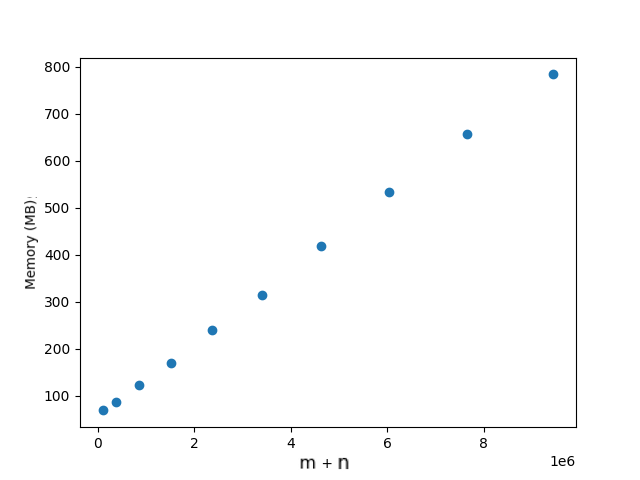}
	\caption{Memory usage of Tarjan algorithm on \textit{Yersinia.wgMLST} dataset.}
	\label{fig:mem_eval}
\end{figure}

We evaluated also the memory requirements for dynamic updating.
We only measured the memory consumption for the \texttt{DELETE} operation because the \texttt{ADD} operation is essentially reduced to an edge removal operation.
Tables~\ref{table:clostridium.Griffiths_mem_cmp} to~\ref{table:Yersiniawgs_mem_cmp} show the memory usage comparison between the ab initio computation and the dynamic updating, averaged over $10$ operations.
In each table, the first column contains the \% of the dataset being considered, the second column presents the memory usage for the ab initio computation, the third column presents the memory usage for the dynamic updating, and the fourth column presents the memory ratio between dynamic updating and ab initio computation.
As an illustrative baseline, Figure~\ref{fig:mem_eval} shows the memory usage for \textit{Yersinia.wgMLST} dataset as an increasing percentage of it added to the computation.
Given these results, we can observe that both ab initio computation and dynamic updating require linear space on the size of the input.
This is also consistent with the results for random graphs presented above.
However the dynamic updating requires three times more memory on average than the ab initio computation, which is expected given that a more complex data structure needs to be managed.
\begin{table}[H]
	\centering
	\caption{Memory usage comparison for the dynamic updating and ab initio computation of an optimal arborescence for the \textit{clostridium.Griffiths} dataset.}
	\label{table:clostridium.Griffiths_mem_cmp}
	\begin{tabular}{c c c c}
		\hline
		\Longstack{Dataset\\\%}  & \Longstack{Ab initio\\(MB)} & \Longstack{Dynamic updating\\(MB)} & \Longstack{Memory ratio} \\
		\hline
		10 & 6.94& 21.05 &  3.03\\ 
		20 & 7.87& 24.14 & 3.07\\ 
		30 &  8.90& 27.73& 3.12\\ 
		40 & 10.54& 32.52& 3.09\\ 
		50 & 12.54& 39.12& 3.12\\ 
		60 & 14.93&  46.76& 3.11 \\ 
		70 & 17.98& 55.34& 3.08\\ 
		80 & 21.42& 65.21& 3.05\\ 
		90 & 26.29& 77.17& 2.94\\ 
		100 & 31.18& 89.37& 2.87\\  \hline
	\end{tabular}
\end{table}
\begin{table}[H]
	\centering
	\caption{Memory usage comparison for the dynamic updating and ab initio computation of an optimal arborescence for the \textit{Moraxella.Achtman7GeneMLST} dataset.}
	\label{table:moraxella_mem_cmp}
	\begin{tabular}{c c c c}
		\hline
		\Longstack{Dataset\\\%}  & \Longstack{Ab initio\\(MB)} & \Longstack{Dynamic updating\\(MB)} & \Longstack{Memory ratio} \\
		\hline
		10 & 7.93& 23.97& 3.02\\ 
		20 & 9.71& 30.16& 3.11\\ 
		30 & 13.17& 40.63& 3.09\\ 
		40 & 17.10& 54.86& 3.21\\ 
		50 & 23.80& 74.08& 3.21\\
		60 & 30.80& 106.20& 3.45\\ 
		70 & 43.92& 133.99& 3.05\\ 
		80 & 49.65& 163.36&  3.29\\ 
		90 & 63.66& 219.69& 3.45\\ 
		100 & 79.70& 263.23& 3.30\\ 
		\hline
	\end{tabular}
\end{table}
\begin{table}[H]
	\centering
		\caption{Memory usage comparison for the dynamic updating and ab initio computation of an optimal arborescence for the \textit{Yersinia.cgMLSTv1} dataset.}
		\label{table:Yersiniacg_mem_cmp}
	\begin{tabular}{c c c c}
		\hline
		\Longstack{Dataset\\\%}  & \Longstack{Ab initio\\(MB)} & \Longstack{Dynamic updating\\(MB)} & \Longstack{Memory ratio} \\
		\hline
		10 & 20.96& 85.10& 4.06\\ 
		20 & 38.98& 142.42& 3.65\\
		30 & 78.79& 240.36& 3.05\\
		40 & 131.39& 415.18& 3.16\\
		50 & 195.35& 565.25& 2.89\\
		60 & 261.86& 841.66& 3.21\\
		70 & 376.26&  1100.50 & 2.92\\
		80 & 465.27& 1374.02 & 2.95\\
		90 & 612.67& 1684.4 & 2.75\\
		100 & 724.01& 1997.67 & 2.76\\
	    \hline
	\end{tabular}
\end{table}

\begin{table}[H]
	\centering
	\caption{Memory usage comparison for the dynamic updating and ab initio computation of an optimal arborescence for the \textit{Yersinia.wgMLST} dataset.}
	\label{table:Yersiniawgs_mem_cmp}
	\begin{tabular}{c c c c}
		\hline
		\Longstack{Dataset\\\%}  & \Longstack{Ab initio\\(MB)} & \Longstack{Dynamic updating\\(MB)} & \Longstack{Memory ratio} \\
		\hline
		10 & 125.32&  415.93 & 3.32\\ 
		20 & 156.31&  513.5 & 3.29\\ 
		30 & 214.65& 680.62 & 3.17\\ 
		40 & 295.80& 908.01 & 3.07\\ 
		50 & 411.26& 1201.33 & 2.92\\ 
		60 & 534.45& 1585.99 & 2.97\\ 
		70 & 714.39& 2056.44 & 2.88\\ 
		80 & 898.64& 2479.41 & 2.76\\ 
		90 & 1098.49& 3026.20 & 2.75\\ 
		100 & 1306.95& 3640.20 & 2.79\\ \hline
	\end{tabular}
\end{table}

\begin{table}[H]
	\centering
	\caption{Memory usage comparison for the dynamic updating and ab initio computation of an optimal arborescence for the \textit{Salmonella.Achtman7GeneMLST} dataset.}
	\label{table:salmonella_mem_cmp}
	\begin{tabular}{c c c c}
		\hline
		\Longstack{Dataset\\\%}  & \Longstack{Ab initio\\(MB)} & \Longstack{Dynamic updating\\(MB)} & \Longstack{Memory ratio} \\
		\hline
		10 & 62.60& 210.45 & 3.36\\ 
		20 & 154.52& 520.83 & 3.37\\ 
		30 & 319.87& 1041.22 & 3.26\\ 
		40 & 575.05& 1772.00 & 5.54\\ 
		50 & 886.04& 2667.96& 3.01\\ 
		60 & 1274.59& 3883.46& 3.04\\ 
		70 & 1754.89& 5242.21 & 2.99\\ 
		80 & 2285.91& 6841.03 & 2.99\\ 
		90 & 2872.80&  8574.12 & 2.99\\ 
		100 & 3435.51& 10482.84 & 3.05\\ \hline
	\end{tabular}
\end{table}

\section{Conclusions}

We provided implementations of Edmonds algorithm and of Tarjan algorithm
for determining optimal arborescences on directed and weighted graphs.  Our
implementation of Tarjan Algorithm incorporates the corrections by Camerini et
al., and it runs in $(m\log n)$ time, where $n$ is the number of vertices of
the graph and $m$ is the number o edges.  We provide also an implementation for
the dynamic updating of optimal arborescences based on the ideas by Pollatos,
Telelis and Zissimopoulos, and that relies on Tarjan algorithm, running in $O(n + ||\rho|| \log |\rho|)$
per update operation and scaling linearly with respect to memory usage.
We highlight the fact that our implementations are generic in the sense that a generic comparator is given as parameter and, hence, we are not restricted to weighted graphs; we can find the optimal arborescence on any graph equipped with a total order on the set of edges.
To our knowledge, our implementation for optimal arborescence problem on dynamic graph is the first one to be publicly available.
The code is available at \url{https://gitlab.com/espadas/optimal-arborescences}.

Experiment evaluation shows that our implementations comply with the expected theoretical bounds.
Moreover, while multiple changes occur in $G$, the dynamic updating is at least twice as faster 
as the ab initio computation, requiring although more memory even if by a constant factor.
Our experimental results corroborate also the results presented by B\"{o}ther et al. and Pollatos et al.

We found one shortcoming regarding the dynamic optimal arborescence, namely the
high dependence between the time needed to recalculate the optimum arborescence
and the affected level of the ATree.  The lower the level, the larger the
number of affected constituents will be.  A prospect to achieve a more
efficient dynamic algorithm could be relying on link-cut trees~\cite{ST83}
which maintains a collection of node-disjoint forests of self-adjusting binary
heaps (splay-trees~\cite{RUSSO20191}) under a sequence of \texttt{LINK} and
\texttt{CUT} operations.  Both operation take $O(\log n)$ time in worst-case.

With respect to the application in the phylogenetic inference context, we
highlight the fact that the proposed implementation for dynamic updates allows
to significantly improve the time required to update phylogenetic patterns as
datasets grow in size. We note also that, due to the use of
heuristics in the  probable optimal tree inference, there are some algorithms
that include a final step for further local optimizations~\cite{zhou2018grapetree}.
Although it may not be always the case, it seems that we can often incorporate
such local optimization as total order over edges.
Given that our implementations assume that such a total order is given as parameter,
such optimizations can be incorporated. The challenge of combining these
techniques to implement classes of local optimizations is also a path for
future work.

\authorcontributions{JE and APF designed and implemented the solution. JE, LMSR, TR and CV conducted the experimental evaluation. CV, APF, LMSR and TR wrote the manuscript. All authors wrote, read and approved the final manuscript.}

\funding{The work reported in this article received funding from Funda\c{c}\~ao para a Ci\^encia e a Tecnologia (FCT) with references UIDB/50021/2020, LA/P/0078/2020 and PTDC/CCI-BIO/29676/2017 (NGPHYLO project), and from European Union’s Horizon 2020 research and innovation program under Grant Agreement No. 951970 (OLISSIPO project). It was  also supported through Instituto Politécnico de Lisboa with project IPL/IDI\&CA2023/PhyloLearn\_ISEL}

\appendixtitles{no}
\begin{adjustwidth}{-\extralength}{0cm}
\reftitle{References}
\bibliography{bib}

\end{adjustwidth}
\end{document}